\documentclass[10pt]{article}
\pdfoutput=1
\usepackage[a4paper, total={19cm, 27cm}]{geometry}
\usepackage{graphicx}
\usepackage{epsf}
\usepackage{amsmath}
\usepackage[normalem]{ulem}
\usepackage{amssymb}

\usepackage{cite}
\usepackage{multirow,tabularx}
\usepackage{appendix}
\usepackage{tikz}
\usepackage{amsmath,amsfonts,amssymb,amsthm,euscript,braket,xcolor}
\newcommand{\be}{\begin{equation}}
\newcommand{\ee}{\end{equation}}

\newcommand{\Rmnum}[1]{\expandafter\@slowromancap\romannumeral #1@}
\newcommand{\bea}{\begin{eqnarray}}
\newcommand{\eea}{\end{eqnarray}}

\numberwithin{equation}{section}

\usepackage[normalem]{ulem}

\usepackage{subfigure}
\begin{document}

\title{\bf Chiral transition in the probe approximation from an Einstein-Maxwell-dilaton gravity model}

\author{\textbf{Hardik Bohra$^{a,b}$}\thanks{bohra.hardik@uky.edu}, \textbf{David Dudal$^{c,d}$}\thanks{david.dudal@kuleuven.be},
\textbf{Ali Hajilou$^{e,f}$}\thanks{ma.hajilou@gmail.com}, \textbf{Subhash Mahapatra$^{a}$}\thanks{mahapatrasub@nitrkl.ac.in}
 \\\\
 \textit{{\small $^a$ Department of Physics and Astronomy, National Institute of Technology Rourkela, Rourkela - 769008, India}}\\
 \textit{{\small $^b$Department of Physics and Astronomy, University of Kentucky, Lexington, KY 40506, USA}}\\
\textit{{\small $^c$ KU Leuven Campus Kortrijk--Kulak, Department of Physics, Etienne Sabbelaan 53 bus 7657,}}\\
\textit{{\small 8500 Kortrijk, Belgium}}\\
\textit{{\small $^d$  Ghent University, Department of Physics and Astronomy, Krijgslaan 281-S9, 9000 Gent, Belgium}}\\
\textit{{\small $^e$ Department of Physics, Shahid Beheshti University G.C., Evin, Tehran 19839, Iran}}\\
\textit{{\small $^f$ School of Particles and Accelerators, Institute for Research in Fundamental Sciences (IPM), Tehran 19395-5531, Iran}}}

\date{}


\maketitle
\abstract{We refine an earlier introduced 5-dimensional gravity solution capable of holographically capturing several qualitative aspects of (lattice) QCD in a strong magnetic background such as the anisotropic behaviour of the string tension, inverse catalysis at the level of the deconfinement transition or sensitivity of the entanglement entropy to the latter. Here, we consistently modify our solution of the considered Einstein-Maxwell-dilaton system to not only overcome an unphysical flattening at large distances in the quark-antiquark potential plaguing earlier work, but also to encapsulate inverse catalysis for the chiral transition in the probe approximation. This brings our dynamical holographic QCD model yet again closer to a stage at which it can be used to predict  magnetic QCD quantities not directly computable via lattice techniques.}

\section{Introduction}
Subjecting QCD to extreme external conditions such as temperature, density and/or electromagnetic fields is not only a matter of formal and theoretically challenging studies, but of direct possible relevance for current particle accelerator driven research programs \cite{Ackermann:2000tr,Bannier:2014bja,Lohner:2012ct}, early universe physics \cite{Grasso:2000wj,Vachaspati:1991nm}, dense neutron stars \cite{Duncan:1992hi}, gravitational waves physics \cite{Ecker:2019xrw}, etc.

Next to the necessary high temperature conditions to liberate quarks from their permanent confinement, (non-central) relativistic heavy ion collisions might also create during the short-lived quark-gluon plasma stage \cite{McLerran:2013hla,Tuchin:2013apa}, a strong magnetic background \cite{Kharzeev:2007jp,Skokov:2009qp,Bzdak:2011yy,Voronyuk:2011jd,Deng:2012pc,Tuchin:2013ie}, another player affecting the QCD phase diagram \cite{Miransky:2015ava,Kharzeev:2012ph,DElia:2010abb}.

Unfortunately, understanding the QCD phase structure under the aforementioned circumstances remains challenging \cite{Stephanov:2007fk}. Analytical approaches are hard because of the strong coupling, and there are always some modelling or truncation artifacts, etc. The simulation-based approach of lattice QCD is a powerful ally, but due to the inherently Euclidean nature of the Monte Carlo setup, the effects of a chemical potential or transport coefficients---related to out-of-equilibrium physics, are conceptually difficult to access because of the infamous sign-problem.
The modern tensor network paradigm does not suffer from this particular conceptual drawback, but as of now seems to be computationally limited to lower-dimensional gauge theories \cite{Banuls:2019bmf}.

Another option---the one we will follow here---is applying the gauge-gravity correspondence rooted in \cite{Maldacena:1997re,Gubser:1998bc,Witten:1998qj}, which has become a key player in the field
of theoretical studies of the strongly coupled quark-gluon plasma. A key modification of the original AdS/CFT vocabulary is that QCD requires a
mass scale/confinement, not available in a conformal setting. This means the original AdS gravity
background is untenable and needs to be replaced by more involved backgrounds reflecting the
fundamental QCD scale, allowing at least for confinement and a massive spectrum built from
the (almost) massless original degrees of freedom. Relevant examples of such AdS/QCD theories
are \cite{Sakai,Erlich:2005qh,wall2}, some examples of the role of anisotropy, as brought in by a magnetic field, in dual gauge theories can be found in e.g.~\cite{Giataganas:2013hwa,Gursoy:2018ydr,Mateos:2011ix,Giataganas:2012zy,Arefeva:2018cli,Dudal:2014jfa,Dudal:2016joz,Rougemont:2014efa,Critelli:2016cvq,
Braga:2018zlu,Braga:2019yeh,Zhou:2020ssi}.

In this work, we will mainly focus on the inverse magnetic catalysis of chiral symmetry breaking, something which was first perceived as unexpected and counterintuitive:  indeed, the earlier papers \cite{Gusynin:1994xp,Gusynin:1994re} gave support to a magnetic field induced catalysis of the chiral condensate, from which naively a larger chiral restoration temperature could be guessed. Nonetheless, the opposite behaviour received convincing lattice evidence from \cite{Bali:2011qj}, see also \cite{Bali:2012zg,Ilgenfritz:2013ara}. Likewise, the deconfinement transition temperature also follows this inverse catalysis behaviour. An incomplete list on (inverse) magnetic catalysis motivated works is \cite{Miransky:2002rp,Gatto:2010pt,Mizher:2010zb,Osipov:2007je,Kashiwa:2011js,Alexandre:2000yf,Fraga:2008um,Fukushima:2012xw,Semenoff:1999xv,Shovkovy:2012zn,Bali:2011qj,Bali:2012zg,
Ilgenfritz:2013ara,Bruckmann:2013oba,Fukushima:2012kc,Ferreira:2014kpa,Mueller:2015fka,Bali:2013esa,Fraga:2012fs,Ayala:2014iba,Ayala:2014gwa,Fraga:2012ev}, with holographic contributions being, for example, \cite{Johnson:2008vna,Callebaut:2013ria,Alam:2012fw,Preis:2010cq,Filev:2010pm,Dudal:2015wfn,Mamo:2015dea,Li:2016gfn,Evans:2016jzo,Bolognesi:2011un,
Ballon-Bayona:2017dvv,Rodrigues:2017cha,Rodrigues:2017iqi,McInnes:2015kec,Gursoy:2017wzz,Gursoy:2016ofp,Dudal:2018rki,Rodrigues:2018pep,Arefeva:2020uec,He:2020fdi,Braga:2020hhs,Braga:2020opg,Rougemont:2015oea,Fang:2019lmd,Fang:2019xbk}.

This paper continues the study laid out in \cite{Bohra:2019ebj}. To model QCD in a magnetic background, we will rely on an exact solution of the Einstein-Maxwell-dilaton (EMD) gravity system,  obtainable via the potential reconstruction method \cite{He:2013qq,Yang:2015aia,Dudal:2017max,Arefeva:2018hyo,Arefeva:2020byn,Dudal:2018ztm,Mahapatra:2018gig,Mahapatra:2019uql,Li:2011hp,Cai:2012xh,Alanen:2009xs,Mahapatra:2020wym}. This solution, depending on a scale function $A(z)$ that can be chosen at will to mimic desired QCD features, incorporates a magnetic field as well as a running dilaton. A short survey of it will be presented in Sect.~2, including the Hawking-Page transition \cite{Hawking} in terms of the magnetic field. Relative to \cite{Bohra:2019ebj}, we will show how a small deformation of the considered form factor $A(z)$ can resolve an unphysical feature in the heavy (static) quark potentials computed in \cite{Bohra:2019ebj} via the holographic Wilson loop prescription \cite{Maldacena:1998im,Brandhuber:1998bs,Rey:1998ik,Andreev:2006ct}, namely that the increasing-linear-in-distance (confining) behaviour is replaced by a constant value at large quark-antiquark separation. As we do not include light dynamical quarks, this flattening behaviour, corresponding to string breaking, should not occur. Here, we will show that this can be averted, restoring the linear increase to arbitrary separation.

In Sect.~3, we will add flavour (quark) matter via a phenomenological probe brane construction and study the occurrence of (inverse) magnetic catalysis at the level of the chiral phase transition. For this we shall implement two different numerical techniques, and use the form-factors introduced in Sec.~2.

We end with a short outlook to further research in Sect.~4.

\section{Survey of the gravitational background}
\subsection{Magnetized Einstein-Maxwell-dilaton gravity}
In order to study the effect of a magnetic field and chemical potential on some features of QCD in the context of holographic QCD (AdS/QCD) models, we rely on a gravity background with two Maxwell fields, a first one dual to the chemical potential (or better said, the neutral baryon number current), and a second one dual to the electromagnetic current in the boundary field theory\footnote{In this paper we will not be interested in studying the effect of a chemical potential, so we will set it equal to zero after this subsection.}. In such $U(1)\times U(1)$ setup, mesons are
charge neutral, so we cannot directly couple electromagnetism to the theory. We will employ the second gauge field just to introduce a (constant)
magnetic field i.e., we have no interest in the associated mesonic fluctuations.

So, the background that we have utilized is five-dimensional EMD gravity \cite{Bohra:2019ebj},
\begin{eqnarray}
S_{EM} =  -\frac{1}{16 \pi G_5} \int \mathrm{d^5}x \sqrt{-g}  \ \left[R - \frac{f_{1}(\phi)}{4}F_{(1) MN}F_{(1)}^{MN}- \frac{f_{2}(\phi)}{4}F_{(2)MN}F_{(2)}^{MN} -\frac{1}{2}\partial_{M}\phi \partial^{M}\phi -V(\phi)\right]\,,
\label{actionEMD}
\end{eqnarray}
where $G_5$ is the Newton constant in five dimensions, $R$ is the Ricci scalar, $\phi$ is the dilaton field, $F_{(1)MN}$ and $F_{(2)MN}$ are the field strength tensors for the two $U(1)$ gauge fields, $f_{1}(\phi)$ and $f_{2}(\phi)$ are the gauge kinetic functions that act as coupling between gauge fields and dilaton field and $V(\phi)$ is the potential of the dilaton field (for more details about this action see \cite{Bohra:2019ebj}).

To obtain the on-shell solutions, the following Ans\"{a}tze have been considered for the metric field $g_{MN}$, dilaton field $\phi$ and electromagnetic field tensors $F_{(i)MN}$:
\begin{eqnarray}
& & ds^2=\frac{L^2 S(z)}{z^2}\biggl[-g(z)dt^2 + \frac{dz^2}{g(z)} + dy_{1}^2+ e^{B^2 z^2} \biggl( dy_{2}^2 + dy_{3}^2 \biggr) \biggr]\,, \nonumber \\
& & \phi=\phi(z), \ \ A_{(1) M}=A_{t}(z)\delta_{M}^{t}, \ \  F_{(2)MN}=B dy_{2}\wedge dy_{3}\,,
\label{ansatze}
\end{eqnarray}
where $L$ is the AdS length scale \footnote{We set $L=1$ in the numerical calculations.}, $S(z)$ is the scale factor, and $g(z)$ is the blackening function. Here, $z$ is the radial coordinate with $z = 0$ at the AdS boundary. This coordinate $z$ runs from the boundary to the horizon at $z=z_h$ for the black hole case or to $z=\infty$ for the thermal-AdS case. Also, the magnetic field $\vec B$ is located parallel to the $y_1$-direction. We remind here that this magnetic field $B$ (mass dimension 1) is actually the 5-dimensional one, which needs a rescaling via the AdS length $L$ to get the physical, 4-dimensional,
boundary magnetic field $\cal B$ (mass dimension 2) \cite{Dudal:2015wfn}, see also \cite{DHoker:2009ixq}.  For our current qualitative purposes, we will keep using the 5-dimensional magnetic field. Notice this is more than just working in units $L=1$, there is still an undetermined dimensionless ratio between the two magnetic fields.

Utilizing the above Ans\"{a}tze eq.~(\ref{ansatze}), imposing suitable boundary conditions and following the procedure outlined in \cite{Bohra:2019ebj}, complete solutions can be expressed in terms of two arbitrary functions, i.e.~the gauge coupling function $f_1(z)$ and the scale function $S(z)$. Doing so, the solution for $A_{t}(z)$ is
\begin{eqnarray}
A_{t} (z) =\frac{\mu}{ \int_0^{z_h} \, d\xi \frac{ \xi e^{-B^2 \xi^2}}{f_{1}(\xi) \sqrt{S(\xi)}} }
    \biggl[ \int_z^{z_h} \, d\xi \frac{ \xi e^{-B^2 \xi^2}}{f_{1}(\xi)
   \sqrt{S(\xi)}} \biggr] = \tilde{\mu} \int_z^{z_h} \, d\xi \frac{ \xi e^{-B^2 \xi^2}}{f_{1}(\xi)
   \sqrt{S(\xi)}} \,,
 \label{Atsol}
\end{eqnarray}
and the solution for gauge coupling function $f_2(z)$
\begin{eqnarray}
f_2(z) = - \frac{ e^{2 B^2 z^2} L^2 S(z)}{z}\biggl[ g(z) \left(4 B^2 z+\frac{3
   S'(z)}{S(z)}-\frac{4}{z}\right)+2 g'(z) \biggr] \,,
\label{f2sol}
\end{eqnarray}
while for the potential $V(z)$
\begin{eqnarray}
V(z) = \frac{g(z)}{L^2} \left(-\frac{9 B^2 z^3 S'(z)}{2 S(z)^2}+\frac{10
   B^2 z^2}{S(z)}-\frac{3 z^2 S'(z)^2}{S(z)^3}+\frac{12 z
   S'(z)}{S(z)^2}+\frac{z^2 \phi '(z)^2}{2 S(z)}-\frac{12}{S(z)}\right)  \nonumber \\
   -\frac{z^4 f_{1}(z)A_{t}'(z)^2}{2 L^4 S(z)^2}+ \frac{g'(z)}{L^2}
   \left(-\frac{B^2 z^3}{S(z)}-\frac{3 z^2 S'(z)}{2 S(z)^2}+\frac{3
   z}{S(z)}\right)  \,.
\label{Vsol}
\end{eqnarray}
Also, via studying the vector meson mass spectrum and deconfinement transition temperature, one can make an educated guess for the form of gauge coupling functions $f_1(z)$ and $S(z)$ i.e. $f_1(z) = \frac{e^{-c z^2 -B^2 z^2}}{\sqrt{S(z)}} $ and $S(z)=e^{2 A(z)}$. Therefore, one can write the solutions for $g(z)$ and $\phi(z)$ more explicitly as
\begin{eqnarray}
g(z) &=& 1 +  \int_0^z \, d\xi \ \xi^3 e^{-B^2 \xi^2 -3A(\xi) } \biggl[ K_{3} + \frac{\tilde{\mu}^2}{2 c L^2} e^{c \xi^2}  \biggr], \nonumber \\
&&\text{with} \ \ \ K_3 = - \frac{ \biggl[1+\frac{\tilde{\mu}^2}{2 c L^2} \int_0^{z_h} \, d\xi \  \xi^3 e^{-B^2 \xi^2-3A(\xi)+c \xi^2}
 \biggr]}{\int_0^{z_h} \, d\xi \ \xi^3 e^{-B^2 \xi^2-3A(\xi)}}  \,.
\label{gsol}\\
\phi(z) &=& \int \, dz \sqrt{-\frac{2}{z} \left(3 z A''(z)-3 z A'(z)^2+6
   A'(z)+2 B^4 z^3+2 B^2 z\right)} +K_5
\label{phisol}
\end{eqnarray}
where we can fix the constant $K_{5}$ by demanding that $\phi |_{z=0}\rightarrow 0$, $A(z)$ is scale factor, and $c$ is a constant that can be fixed as $c=1.16~\text{GeV}^{2}$, see \cite{Bohra:2019ebj,Dudal:2017max,Yang:2015aia}. Also, to obtain the critical Hawking-Page (deconfinement) transition temperature $T_{crit}$ as a function of magnetic field, we need the black hole temperature and entropy that are
\begin{eqnarray}
T = - \frac{z_h^3 \ e^{-3 A\left(z_h\right)-B^2 z_h^2}}{4 \pi} \biggl[K_3 + \frac{\tilde{\mu}^2}{2 c L^2} e^{c z_{h}^2} \biggr] ,  \  \  \ S = \frac{e^{B^2 z_{h}^{2}+3A(z_h)}}{4 z_{h}^3}.
\label{phicase11}
\end{eqnarray}
It is important to stress that the eqs.~(\ref{Atsol}), (\ref{f2sol}), (\ref{Vsol}), (\ref{gsol}) and (\ref{phisol}) are a complete solution for magnetized EMD gravity, just depending on the form factor choice $A(z)$. The consistency of this potential reconstruction approach was thoroughly addressed in our previous paper \cite{Bohra:2019ebj}, including the compatibility of the on-shell potential with the Gubser stability criterion \cite{Gubser:2000nd}, next to the almost independence on the external parameters $z_h$
(or $T$), $B$ and $\mu$ of the (on-shell) potential $V(z)$.

In the rest of the paper, we will first shortly reconsider our original choice for the  form factor, as used in \cite{Bohra:2019ebj}, and then introduce a slightly modified form factor to overcome the unphysical string breaking. Afterwards, we discuss the chiral phase transition in Sect.~3.

\subsection{The original form factor $A(z)=A_1(z)=-a z^2$}
Similar to \cite{Dudal:2017max,Bohra:2019ebj} the first case for scale factor that we have used is $A(z)=A_1(z)=-a z^2$ where $a$ can be fixed by matching to the deconfinement temperature obtained from lattice QCD at $B=0$, yielding $a=0.15~\text{GeV}^{2}$. Utilizing this form factor, the dilaton field $\phi(z)$ is
\begin{eqnarray}
\phi(z) = \frac{\left(9 a-B^2\right) \log \left(\sqrt{6
   a^2-B^4} \sqrt{6 a^2 z^2+9 a - B^4 z^2 -B^2}+6 a^2 z - B^4 z
   \right)}{\sqrt{6 a^2-B^4}} \nonumber \\
   +z \sqrt{6 a^2 z^2+9 a - B^2 \left(B^2
   z^2+1\right)} - \frac{\left(9 a-B^2\right) \log \left(\sqrt{9 a-B^2} \sqrt{6
   a^2-B^4}\right)}{\sqrt{6 a^2-B^4}} \,.
\label{phifinal1}
\end{eqnarray}
From eq.~(\ref{phifinal1}) we see that there is maximal value of the magnetic field for which our system is physical. Indeed, the dilaton field should be real-valued and this will be satisfied only if $B^4\leq B_c^4=6a^2$. So, when we are working with this form factor $A_1(z)$, the largest attainable value of magnetic field is $B_c \simeq 0.61~\text{GeV}$.

\begin{figure}[h!]
\centering
\includegraphics[width=2.7in,height=1.9in]{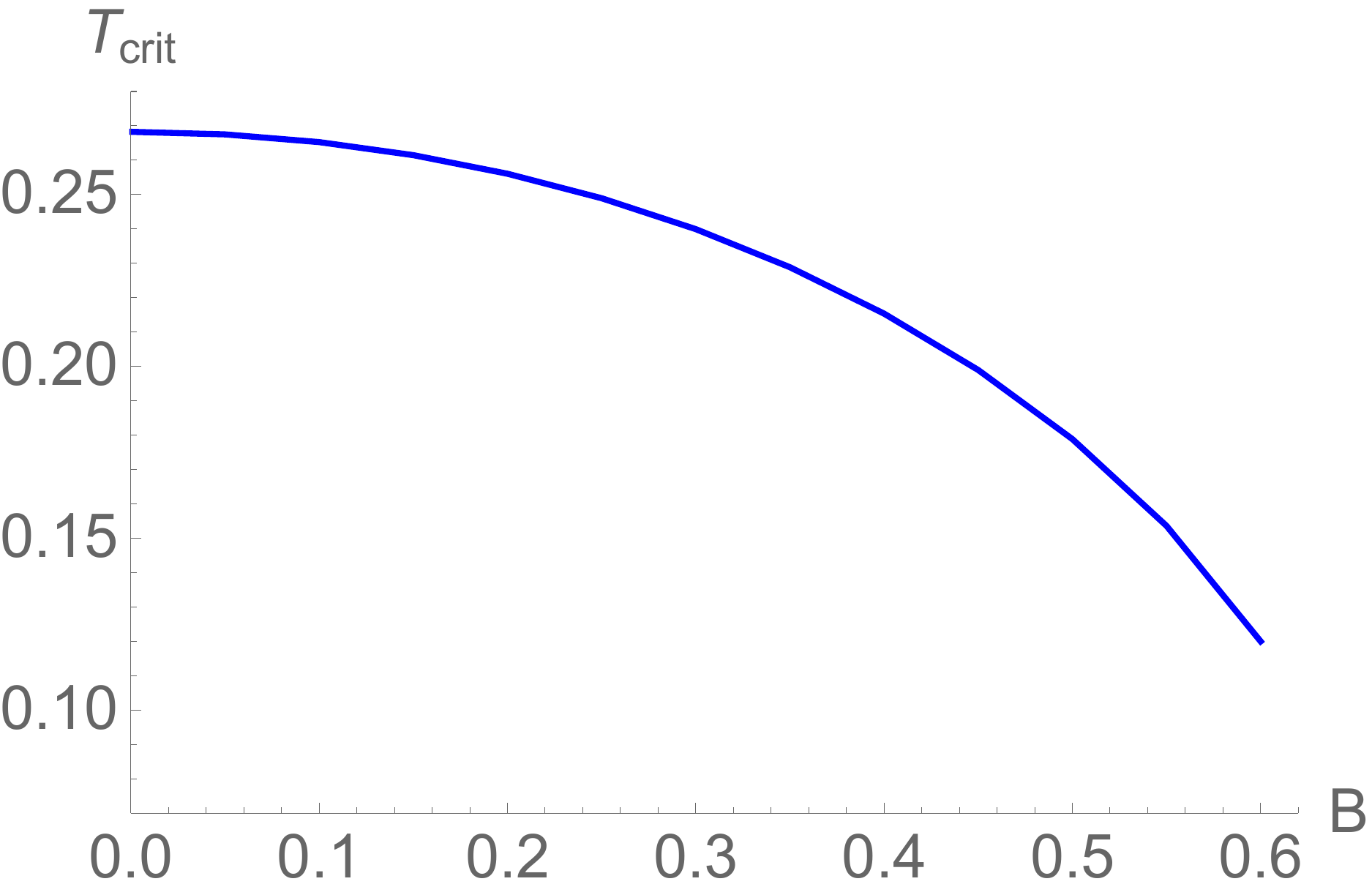}
\caption{ \small  Deconfinement transition temperature in terms of magnetic field for the case $A_1(z)=-a z^2$. Here we set $\mu=0$. In units \text{GeV}.}
\label{tem1}
\end{figure}

In our previous paper \cite{Bohra:2019ebj}, we already showed that the magnetic field induced an inverse magnetic catalysis behaviour for the deconfinement transition temperature, the situation is summarized in Figure~\ref{tem1}.
\subsection{A new form factor: $A(z)=A_2(z)=-a z^2 - d\,B^2 z^5$}
We will now construct a new scale factor that respects two important conditions:
\begin{enumerate}
\item a real-valued dilaton field, preferably also for larger values of the magnetic field.
\item  the free energy of a connected string attached to a boundary quark-antiquark pair should be smaller than the disconnected one\footnote{The connected solution is a $\cup$-shape configuration starting at the quark-antiquark pair living on the boundary ($z=0$) and extends into the bulk. On the other hand, the disconnected solution is a configuration of two lines extending from the boundary to the horizon \cite{Dudal:2017max,Bohra:2019ebj}.} to assure a confined quark-antiquark pair in the dual boundary theory, at least when the metric is of the thermal-AdS type (no black hole), irrespective of the separation length between quark and antiquark. In \cite{Bohra:2019ebj}, we found that the linearly increasing interquark potential becomes flattened at large separation when the scale factor $A_1(z)$ is employed.
\end{enumerate}

Here, we introduce the new scale factor $A(z)=A_2(z)=-a z^2 - d\,B^2 z^5$, still with the parameter $a$ as before, i.e.~$a=0.15~\text{GeV}^{2}$ as this was determined via comparison with $B=0$ lattice data. For what concerns the value of the extra parameter $d$, for every positive value and as long as the magnetic field is not too large (the concrete maximal value depending on $d$), the dilaton field is real, thence satisfying the first condition. \footnote{The almost independent behaviour of the dilaton potential on the parameters $T$ and $B$ for this new form factor is discussed in Appendix A. }

More restrictions will come from the second condition, let us therefore look at the disconnected free energy of the quark-antiquark \cite{Dudal:2017max,Bohra:2019ebj},
\begin{eqnarray}
{\cal {F}}_{discon}= \frac{L^2}{\pi \ell ^{2}_{s}} \int_\epsilon^{z_h=\infty} dz \frac{e^{2A_{s}(z)}}{z^2}
\label{discon}
\end{eqnarray}
where $\epsilon$ corresponds to a UV cut-off in boundary theory \cite{Ewerz:2016zsx} and the $A_{s}(z)=A_2(z)+\sqrt{\frac{1}{6}}~\phi(z)$ the form factor converted to the string frame, with  $\phi(z)$ given in eq.~(\ref{phisol}). $\ell_s$ is the open string length, which is related to the open string constant as $T_s=1/2\pi \ell^{2}_s$. For the new form factor $A_2(z)$, the disconnected free energy always diverges. Indeed, we expand $\phi(z)$ and $A_s(z)$ in the IR (viz.~at large $z$) and then plug this into eq.~(\ref{discon}),
\begin{eqnarray}
{\cal {F}}_{discon}= \frac{L^2}{\pi \ell ^{2}_{s}} \int^{\infty} dz \frac{e^{2A_{s}(z)}}{z^2}= \frac{L^2}{\pi \ell ^{2}_{s}} \int^{\infty} dz\left[z^4 + \frac{2B^2 z^3}{15d}+\ldots\right]~,
\label{discon2}
\end{eqnarray}
where at $z=\infty$ clearly the disconnected free energy will diverge. So, the ($\epsilon$-regularized) connected free energy will always be smaller than the disconnected one, implying that the quark-antiquark pair will always enjoy confinement.

\begin{figure}[h!]
\begin{minipage}[b]{0.5\linewidth}
\centering
\includegraphics[width=2.8in,height=2.1in]{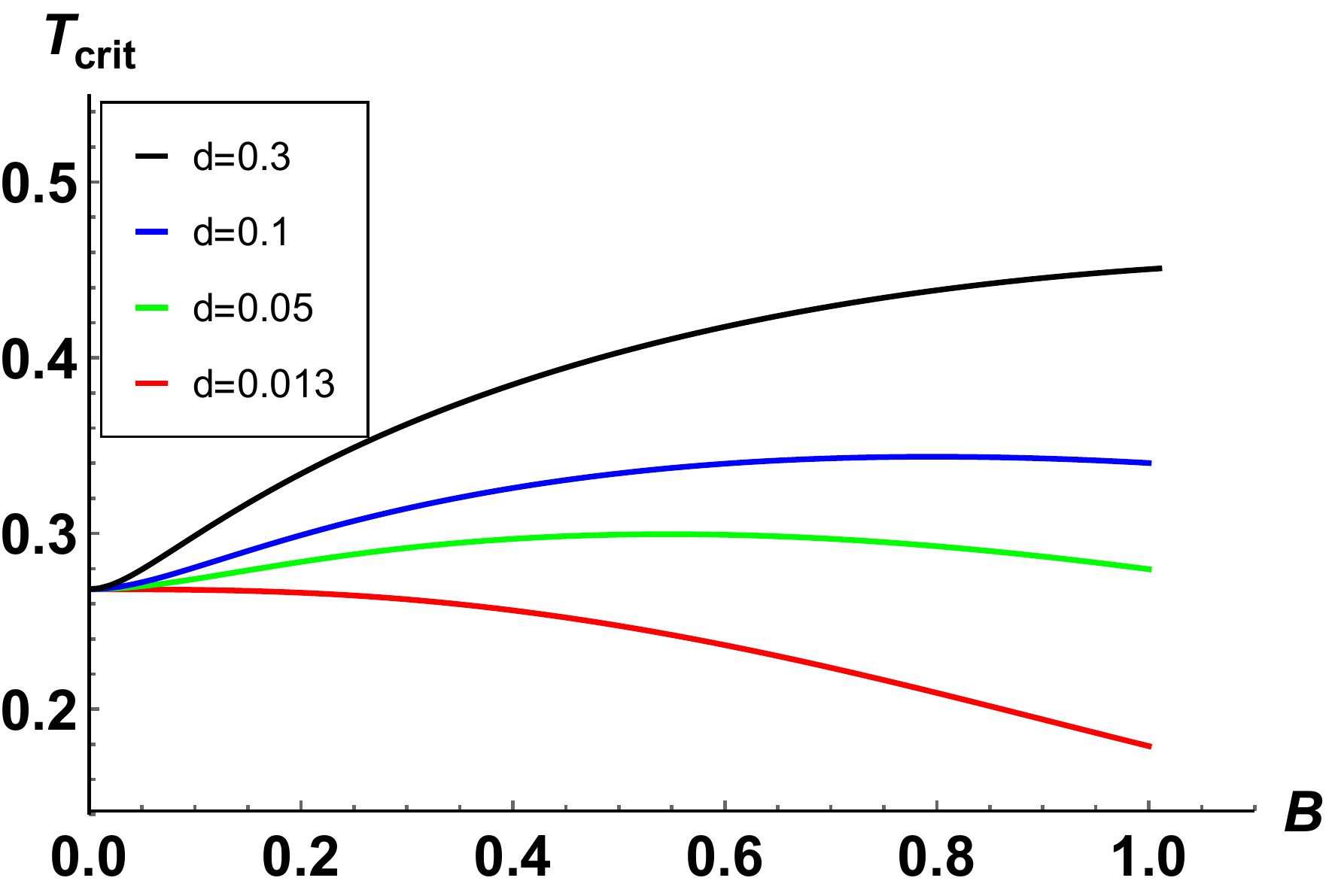}
\caption{ \small Deconfinement transition temperature in terms of magnetic field for the case $A(z)=-a z^2 - d\,B^2 z^5$. Here we set $\mu=0$. In units \text{GeV}.}
\label{thp2}
\end{minipage}
\hspace{0.4cm}
\begin{minipage}[b]{0.5\linewidth}
\centering
\includegraphics[width=2.8in,height=2.1in]{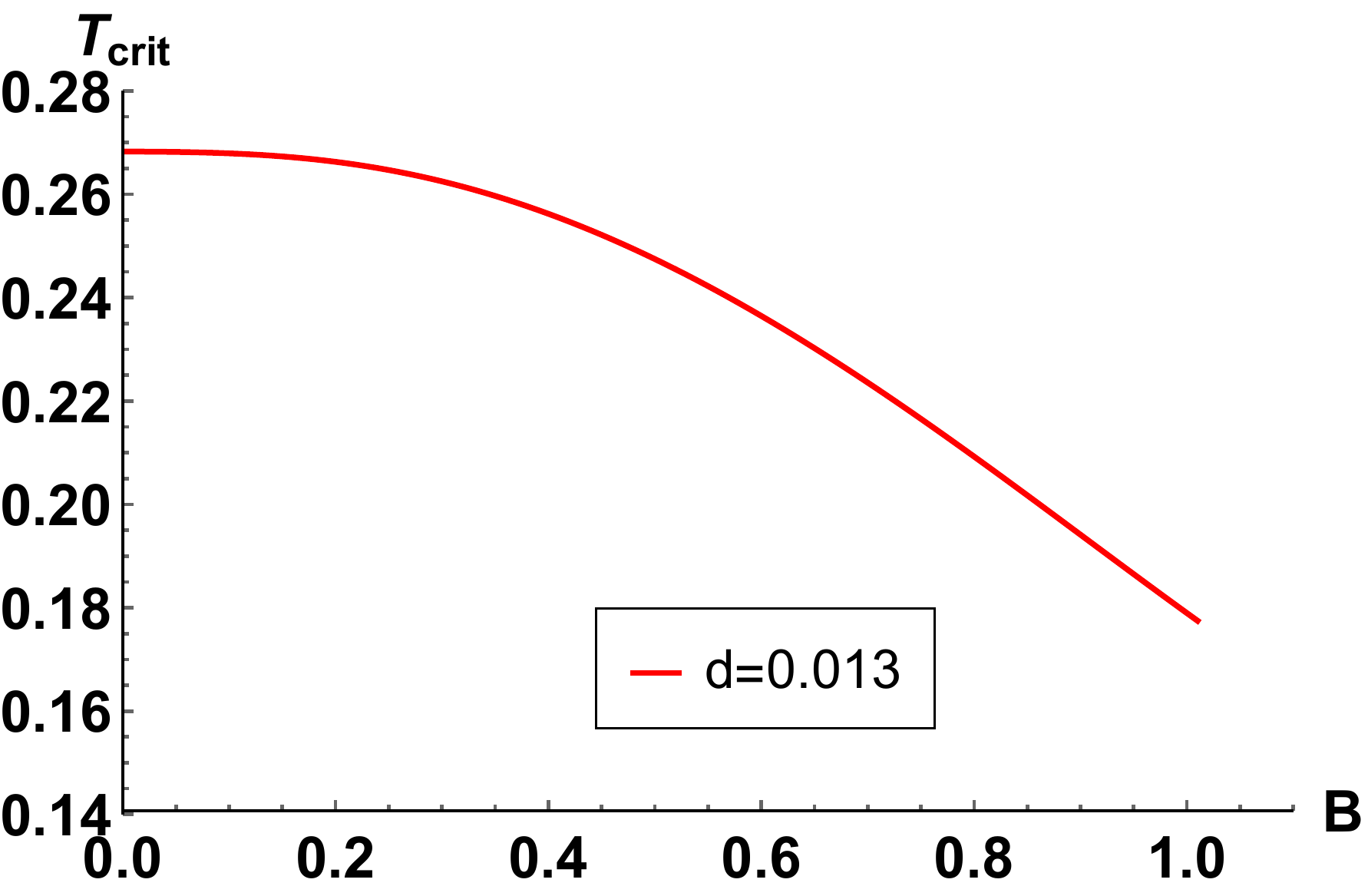}
\caption{\small Deconfinement transition temperature in terms of magnetic field for the case $A(z)=-a z^2 - d\,B^2 z^5$. Here we set $\mu=0$. In units \text{GeV}.}
\label{thp3}
\end{minipage}
\end{figure}

Let us investigate now in a bit more detail the eventual choice of $d>0$, inasmuch as the influence $d$ has on the deconfinement phase transition. We set out the deconfinement transition  temperature $T_{crit}$ in terms of the magnetic field $B$ for different values of $d>0$ in Figure~\ref{thp2}. We observe that inverse magnetic catalysis is persistent just for sufficiently small values of $d$, i.e.~$d\leq 0.013~\text{GeV}^{3}$. We will from now on consider $d=0.013~\text{GeV}^{3}$ to have the largest attainable value for the magnetic field, being $B_c \simeq 1.02~\text{GeV}$. The corresponding confinement-deconfinement phase diagram is displayed in Figure~\ref{thp3}, to be compared with Figure~\ref{tem1} which is the $d=0$ case.

\begin{figure}[h!]
\begin{minipage}[b]{0.5\linewidth}
\centering
\includegraphics[width=2.8in,height=2.0in]{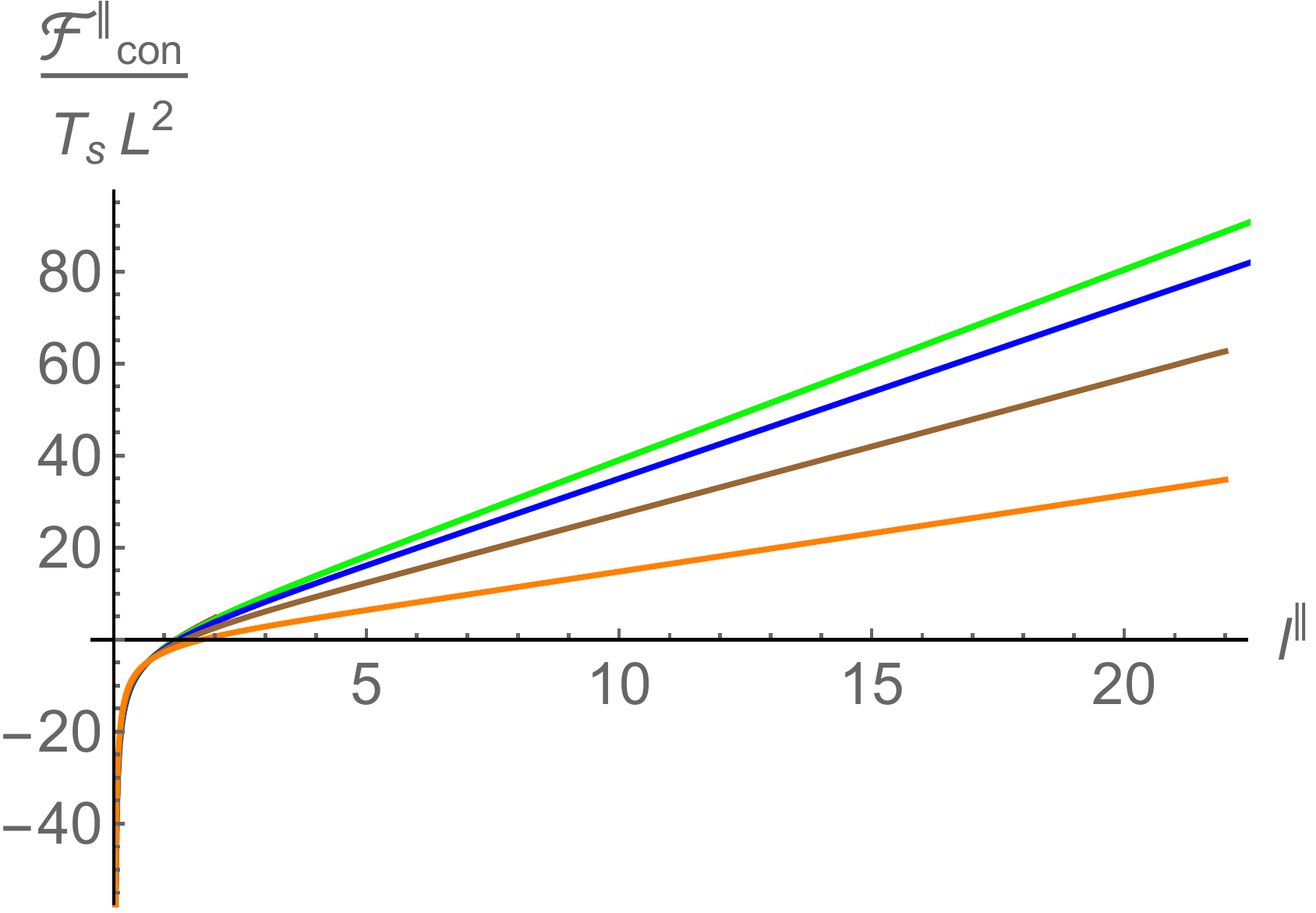}
\caption{ \small The connected free energy ${\cal F}^{\|}_{con}$ as a function of separation length $\ell^{\|}$ in the thermal-AdS background for the case when the Wilson loop is parallel to $\vec B$.  Here $\mu=0$, and red, green, blue, brown and orange curves correspond to $B=0, 0.2, 0.4, 0.6$ and $0.8$ respectively. In units \text{GeV}.}
\label{sigmapara10}
\end{minipage}
\hspace{0.4cm}
\begin{minipage}[b]{0.5\linewidth}
\centering
\includegraphics[width=2.8in,height=2.3in]{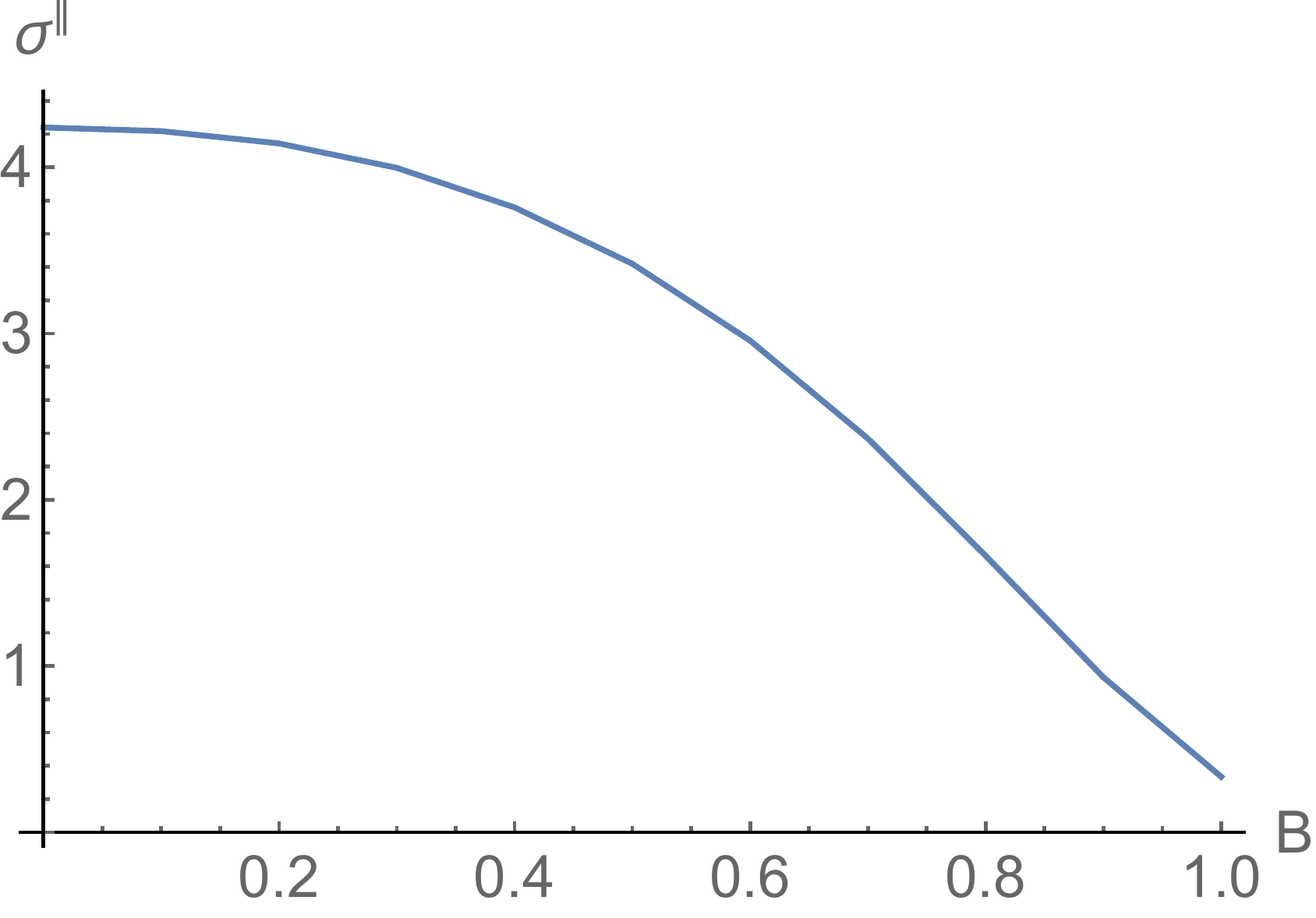}
\caption{\small The string tension in the parallel direction ($\sigma_s^{\|}$) as a function of $B$ in the thermal-AdS background with $\mu=0$. In units \text{GeV}.}
\label{sigmaperp10b}
\end{minipage}
\end{figure}

\begin{figure}[h!]
\begin{minipage}[b]{0.5\linewidth}
\centering
\includegraphics[width=2.8in,height=2.0in]{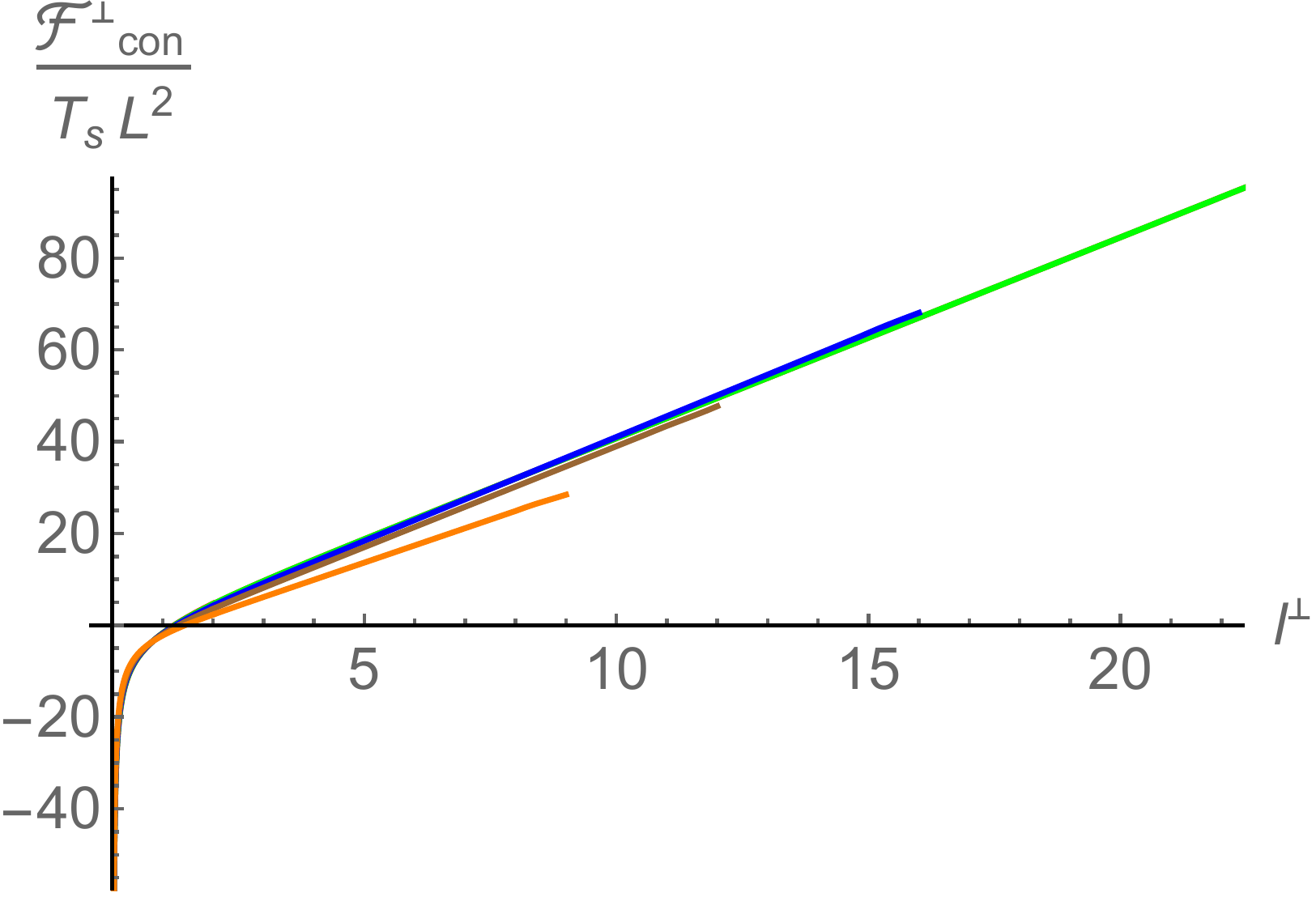}
\caption{ \small  The connected free energy ${\cal F}^{\perp}_{con}$ as a function of separation length $\ell^{\perp}$ in the thermal-AdS background for the case when the Wilson loop is perpendicular to $\vec B$. Here $\mu=0$, and red, green, blue, brown and orange curves correspond to $B=0, 0.2, 0.4, 0.6$ and $0.8$ respectively. In units \text{GeV}.}
\label{sigmapara10c}
\end{minipage}
\hspace{0.4cm}
\begin{minipage}[b]{0.5\linewidth}
\centering
\includegraphics[width=2.8in,height=2.3in]{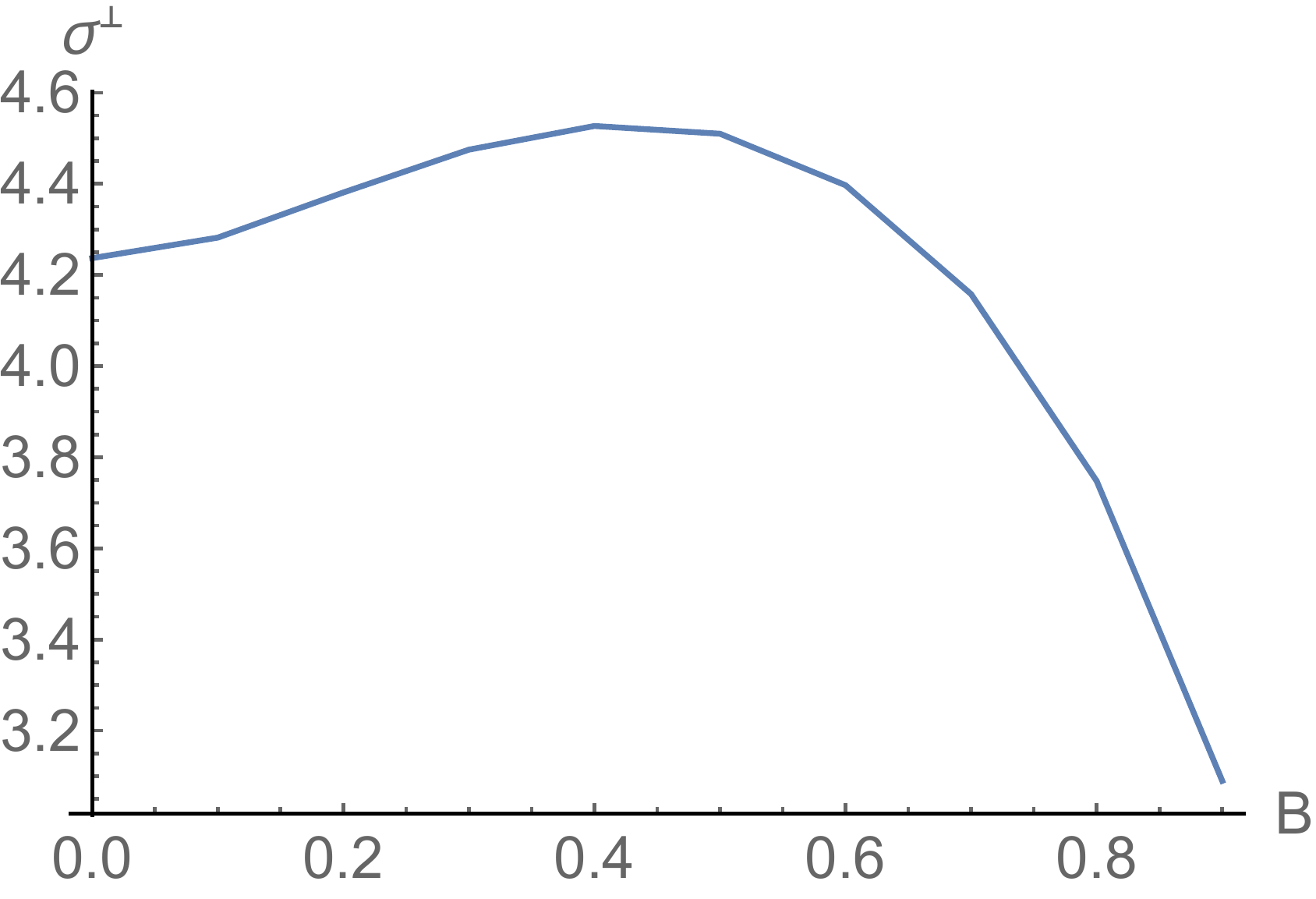}
\caption{\small The string tension in the perpendicular direction $\sigma_s^{\perp}$ as a function of $B$ in the thermal-AdS background with $\mu=0$. In units \text{GeV}.}
\label{sigmaperp10d}
\end{minipage}
\end{figure}

Finally, we delve a bit deeper into the free energy of the quark-antiquark pair, see Figs.~\ref{sigmapara10} and \ref{sigmapara10c} from which it is clear that the flattening obtained in \cite{Bohra:2019ebj} is now avoided. We have determined the string tensions for both parallel and perpendicular orientation of the Wilson loop\footnote{That is, the $q, \bar{q}$ pair is either oriented parallel or perpendicular to the applied magnetic field.}, shown in Figs.~\ref{sigmapara10} and \ref{sigmaperp10d}. Our results for the associated QCD string tensions with this new form factor $A_2(z)$ are compatible with lattice results for smaller values of the magnetic field that were reported in \cite{Bonati:2014ksa,Bonati:2016kxj}, see also \cite{Chernodub:2010bi,Simonov:2015yka} for other approaches: a weaker confinement for the parallel orientation and a stronger one for the perpendicular case. At larger $B$, we also find that the perpendicular string tension starts to decrease again, something not really visible from \cite{Bonati:2014ksa,Bonati:2016kxj}. This being said, unlike these lattice works, we do not have $(2+1)$ dynamical quark flavours in our model. 


\section{Chiral phase transition}
\label{chirals}
In this section, we will investigate the chiral sector of the dual boundary theory and, in particular, investigate the behaviour of chiral condensate and the corresponding chiral critical temperature as a function of the magnetic field. The holographic action relevant for investigating the chiral properties of the boundary QCD theory will be taken as \cite{Colangelo:2011sr,Dudal:2015wfn}\footnote{For more discussion about the action, see Appendix C.},
\begin{eqnarray}
S_{chiral} =  \frac{N_c}{16 \pi^2} \int \mathrm{d^5}x \sqrt{-g} \ e^{-\phi} \ \text{Tr}\left[|DX|^2-m_5^2|X|^2-\frac{f_2(\phi)}{3}(F_{L}^2+F_R^2)\right]\,.
\label{actionchiral}
\end{eqnarray}
Here, $X$ is a $N_f\times N_f$ matrix-valued complex field which is in the bifundamental representation of $SU(N_f)_L\times SU(N_f)_R$. In the AdS/CFT terminology, the field $X^{\alpha,\beta}$ is dual to the quark field operator $\braket{\bar{\psi}^{\alpha}\psi^\beta}$, with $\alpha, \beta$ being the flavour indices, and it is associated with the chiral symmetry breaking on the dual boundary side. $m_5^2$ is the mass of the field $X$ and in this work we will consider $m_5^2 L^2=-3$. $F_{L,R}$ are the field strength tensors for the two (left and right) gauge fields $A_{L,R}$.  The covariant derivative of the chiral field is defined as $D_\mu X=\partial_\mu X- i A_{L,\mu}X+ i X A_{R,\mu}$, which makes the above ``chiral action'' invariant under $SU(N_f)_L\times SU(N_f)_R$ gauge transformations.

A word about the coupling constants. The comparison between the gauge sectors of \eqref{actionchiral} and \eqref{actionEMD} in principle implicitly fixes the Newton constant $G_5$. In future work, we will see how our choice of $f_2(\phi)$ (or better said self-consistently determined solution \eqref{f2sol}) is related to the QCD OPE result for the vector current correlation function, the standard way to fix by hand this gauge coupling constant \cite{Erlich:2005qh}. For the time being, in the present paper we borrowed the prefactor of $X$-sector from \cite{Colangelo:2008us}, where it was matched upon the QCD OPE result for the scalar meson correlator.

For simplicity, following \cite{Dudal:2015wfn,Colangelo:2011sr}, we will work in the approximation of degenerate flavours and consider the field $X$ to be proportional to the identity matrix in flavour
space, i.e.~$X(z,x^\mu)=X_0(z)\textbf{1}_{N_f}e^{i\pi(z,x^\mu)}$, where $X_0(z)$ is the component independent of the boundary directions and $\pi(z,x^\mu)$ represents the chiral field. In this approximation, our main quantity of interest---the chiral condensate---becomes proportional to the quark field operator $\braket{\bar{\psi}^{\alpha}\psi^\alpha}$.  The condensate, therefore, can be extracted by solving the $X$-field equation of motion.

Before we move on to investigate the quark condensate for the different form factors case by case, it is important to point out that there are effectively two ways by which the magnetic field enters in the chiral action: (i) through the background metric, and (ii) via the covariant derivative of $X$. The latter contribution, however, vanishes identically as the magnetic field is introduced into the (diagonal) vector part of the flavour gauge group $A_L=A_R$, for which
\begin{eqnarray}
D_\mu X\to \partial_\mu X \,.
\label{covDchiral}
\end{eqnarray}
Therefore, the only way the magnetic field can influence the quark condensate and chiral critical temperature is through its explicit presence in the background metric.  The sheer importance of this observation can even be more appreciated from the recent work \cite{Ballon-Bayona:2020xtf} where, also in the probe brane approximation whilst using a tachyon condensation-based description of chiral symmetry breaking in the $B=0$ background metric, magnetic catalysis rather than its inverse version was found. To be more precise, our EMD model is \emph{mimicking} QCD in a magnetic background, but we should not forget that we are using a dual (i.e.~probe) version of quenched QCD, that is, no dynamical quarks. As gluons can only couple to the magnetic field through the charged quarks, strictly speaking there should be no dependence of gluon-dominated quantum physics (as confinement-deconfinement, dually encoded in the gravity background) on the magnetic field in a quenched approximation. Constructing a fully dual background, including the flavour brane back reaction, is however highly nontrivial and leads to more intricate, $B$-dependent modeling, such as the V-QCD based ones \cite{Jarvinen:2011qe,Gursoy:2016ofp,Gursoy:2017wzz}. Needless to say, as there is no top-down derivation of ``standard'' QCD from string theory, there is certainly none for magnetized QCD. So, one is always condemned to some level of modelling in QCD-like features \cite{Rougemont:2015oea,Gubser:2008ny}.

At the same time, the black hole metric also allows investigating the temperature-dependent profile of the quark condensate.

Now, a word about the numerical procedure for extracting the chiral information from eq.~(\ref{actionchiral}) is in order. Using eq.~(\ref{covDchiral}), the equation of motion of the $X$-field is given by
\begin{eqnarray}
X_{0}''(z)+X_{0}'(z)\biggl(-\frac{3}{z}+2 B^2 z + \frac{g'(z)}{g(z)} + 3 A'(z) - \phi'(z)  \biggr) + \frac{3 e^{2A(z)}X_0(z)}{z^2 g(z)} =0 \,.
\label{chiraleq}
\end{eqnarray}
The solution to this equation of motion will depend on the confined-deconfined background geometries as well as the form factor $A(z)$. Unfortunately, to the best of our knowledge, the equation is not solvable analytically even for the simplest form factor. However, it can be straightforwardly solved numerically. We employ two different numerical shooting techniques. With the first method, we numerically integrate eq.~(\ref{chiraleq}) from the horizon to the asymptotic boundary and then extract the boundary information using the standard gauge-gravity dictionary. In particular, according to this dictionary, the leading term of the boundary expansion of $X$ starts with the bare quark mass $m_q$ (set by hand) as the lowest order coefficient, whereas the sub-leading term contains the information about the chiral condensate. Therefore, by fixing the bare quark mass by hand, we can integrate eq.~(\ref{chiraleq}) and numerically obtain the chiral condensate. With the second method, we do the opposite and shoot from the AdS boundary until a normalizable solution for the chiral condensate is found, see Appendix D. Needless to say, both these numerical techniques render the same answer.

\subsection{Using the form factor $A(z)=A_1(z)=-a z^2$}
Let us first evaluate the chiral condensate for the simplest case $A(z)=A_{1}(z)=-az^2$. Since the analytic results for the background metric are explicitly known for $A_{1}(z)=-a z^2$, this case will, therefore, allow us to showcase the numerical routine through which we can extract results for the chiral condensate. The results for more complicated form factors can be obtained analogously.

Let us first consider the ultraviolet, near boundary expansion of field $X(z)$ \footnote{We will drop the subscript in $X_0(z)$ from here onwards.}. This is needed for the calculations of physical observables in the dual field theory side. The near boundary expansion of $X$ is given by \footnote{This expansion is valid for both confined and deconfined geometries.},
\begin{eqnarray}
X(z)= m_q z + m_q b_1  z^2 +  \sigma z^3 +  m_q b_2 z^3 \ln{\sqrt{a}z} + \mathcal{O}(z^4) \,,
\label{Xnearzhexp}
\end{eqnarray}
where $m_q$ is the bare quark mass, $b_1=-2\sqrt{9a-B^2}$, and $b_2=-30a+3B^2$. The coefficient $\sigma$ is related to the actual quark condensate $\braket{\bar{\psi}\psi}$ (see Appendix B for more details).  From the boundary expansion \eqref{Xnearzhexp} and the equation of motion \eqref{chiraleq}, it can be easily shown that $\sigma$ scales linearly in $m_q$. This is evidently a non-QCD-like feature (no chiral symmetry breaking in the chiral limit), directly due to the linearity of the equation \eqref{chiraleq}. For now, we will keep using this original chiral setup of e.g.~\cite{Erlich:2005qh,Colangelo:2011sr}.  To produce our plots, we will always set $m_q= 1$~GeV for illustrative purposes.

Since $\sigma$ is a temperature and magnetic field dependent quantity \footnote{The coefficient $\sigma$ here should not be confused with the string tension.}, it gives a temperature and magnetic field dependent profile for the quark condensate $\braket{\bar{\psi}\psi}$. Here, we will use the thermal-AdS background to calculate $\sigma(B,T=0)$ in the confined phase, whereas the black hole background will be used to calculate $\sigma(B,T)$ in the deconfined phase. It should be noted that $\braket{\bar{\psi}\psi}$ actually contains UV divergences.  We are pragmatic here and will rely on a minimal subtraction to only remove the pole parts. The very same strategy was also applied to renormalize the free energies entering the computations in Sect.~2 \cite{Bohra:2019ebj}, based on \cite{Ewerz:2016zsx}. The necessary counterterms in the chiral action for this specific renormalization scheme could also be obtained from a dedicated holographic renormalization analysis as in e.g.~\cite{Ballon-Bayona:2020xtf}.

Here, we will also employ this minimum subtraction renormalization scheme and compare our holographic results with real QCD. Once we adopt this renormalization procedure, the quark condensate $\braket{\bar{\psi}\psi}$ is then related to coefficient $\sigma$ in the following way (see Appendix B for more details),
\begin{eqnarray}
& & \braket{\bar{\psi}\psi}_{B,T} = \frac{N_c}{2 \pi^2} \sigma(B,T)+ \frac{N_c m_q}{8  \pi^2} \biggl( - 18 B^2 + 165 a \biggr)  \,.
\label{psibarpsi3}
\end{eqnarray}

Let us now also briefly discuss the infrared (near horizon) expansion of $X$. Near the horizon, the field $X$ is considered to be smooth and we can assume the following Taylor expansion,
\begin{eqnarray}
X(z)= A_0 + B_0 (z-z_h) + \mathcal{O}(z-z_h)^2 \,.
\label{Xnearzhhor1}
\end{eqnarray}
On substituting eq.~(\ref{Xnearzhhor1}) into eq.~(\ref{chiraleq}) and expanding around the horizon, we get
\begin{eqnarray}
 B_0 = \left(\frac{3 e^{5 A\left(z_h\right)+B^2 z_h^2} \int_0^{z_h} \xi ^3 e^{-3 A(\xi )-B^2 \xi ^2} \, d\xi
   }{z_h^5}-\frac{1}{z_h} \right) A_0 \,.
\label{Xnearzhhor2}
\end{eqnarray}
Therefore, near the horizon $X$ behaves as
\begin{eqnarray}
X = A_0 \biggl[1 + \frac{e^{-5 a z_h^2} \left(2 e^{5 a z_h^2} \left(B^2-3 a\right)^2 z_h^4 +e^{3 a z_h^2} \left(3
   \left(B^2-3 a\right) z_h^2+3\right)-3 e^{B^2 z_h^2}\right)}{2 \left(B^2-3 a\right)^2 z_h^5} (z-zh)
    + \mathcal{O}(z-z_h)^2 \biggr] \,.
\label{Xnearzhhor3}
\end{eqnarray}
Hence, for a fixed $a$, $z_h$ and $B$, there is one independent parameter $X(z_h)$ at the horizon, viz.~$A_0$. This independent parameter can be used to construct an initial value problem for the $X$-field.  The chiral condensate can then be obtained by numerically integrating the $X$-equation of motion from the horizon to the boundary and imposing the boundary condition (\ref{Xnearzhexp}). In particular, integrating out from the horizon to boundary gives a map $X(z_h) \mapsto \sigma$, and this map reduces to a one-parameter family of solutions for each value of $m_q$, $a$, $z_h$ and $B$.

\begin{figure}[h!]
\centering
\includegraphics[width=2.8in,height=2.3in]{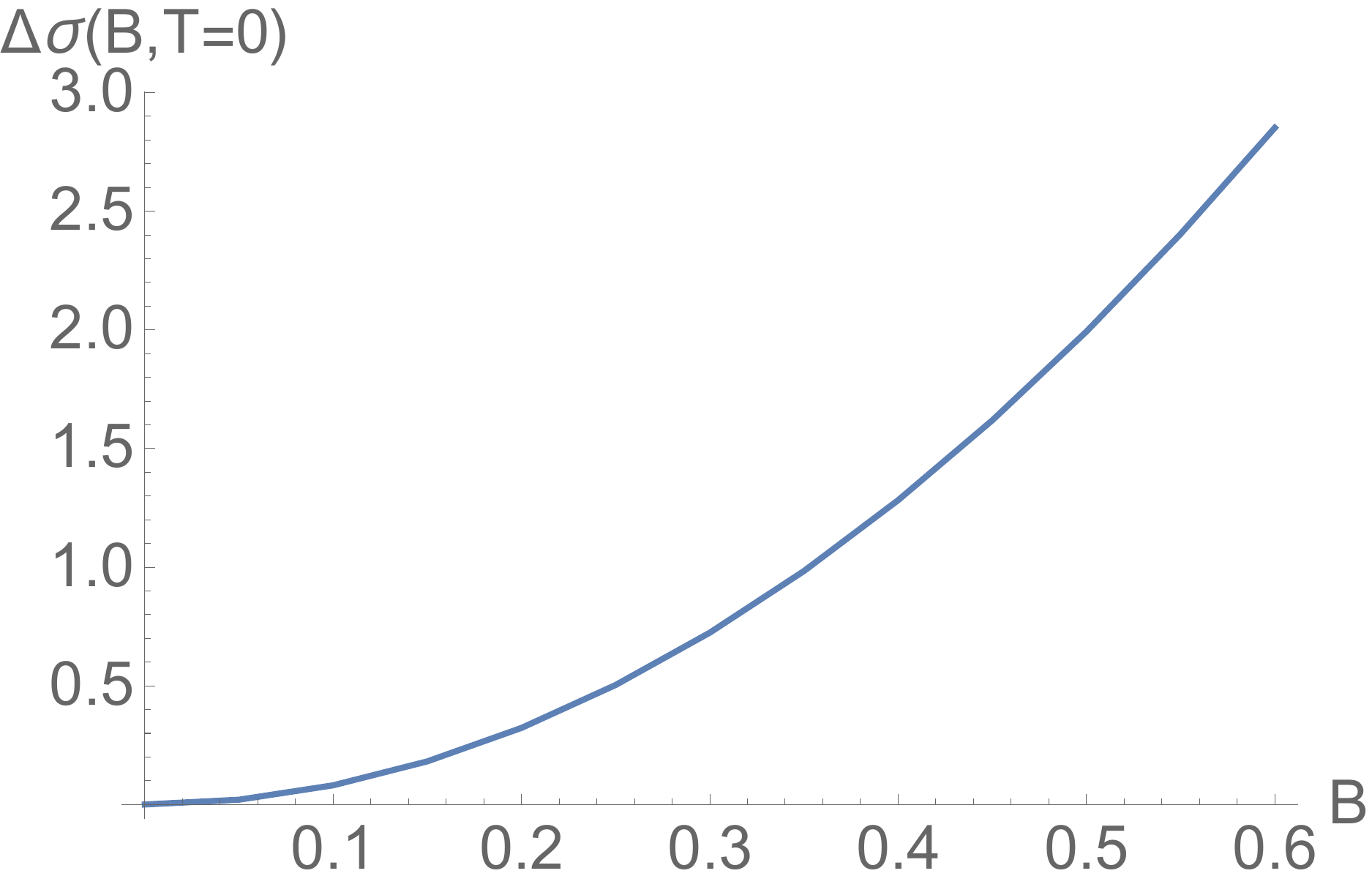}
\caption{\small $\Delta\sigma(B,T=0)=\sigma(B,T=0)-\sigma(B=0,T=0)$ as a function of magnetic field $B$ in the confined phase for the case $A_1(z)=-a z^2$.  Here a quark mass $m_q=1.0$~\text{GeV} is used.}
\label{BvsSigmaMq1confinedalpha1case1}
\end{figure}
\begin{figure}[h!]
\begin{minipage}[b]{0.5\linewidth}
\centering
\includegraphics[width=2.8in,height=2.0in]{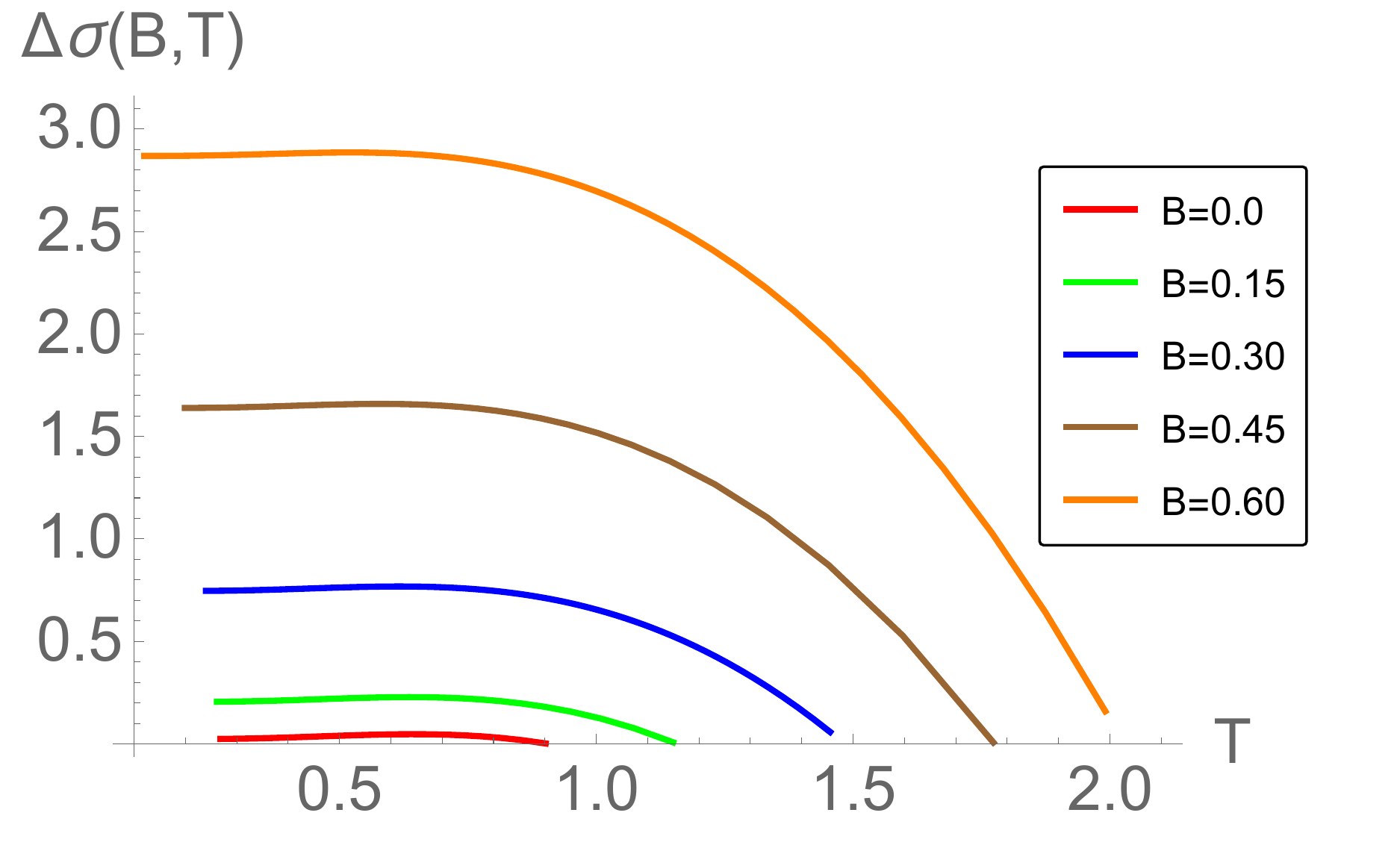}
\caption{ \small $\Delta\sigma(B,T)=\sigma(B,T)-\sigma(B=0,T=0)$ as a function of temperature $T$ in the deconfined phase for the case $A_1(z)=-a z^2$ for different values of the magnetic field.  Here a quark mass $m_q=1.0$~\text{GeV} is used. In units \text{GeV}.}
\label{TvsSigmavsBMq1deconfinedalpha1case1}
\end{minipage}
\hspace{0.4cm}
\begin{minipage}[b]{0.5\linewidth}
\centering
\includegraphics[width=2.8in,height=2.3in]{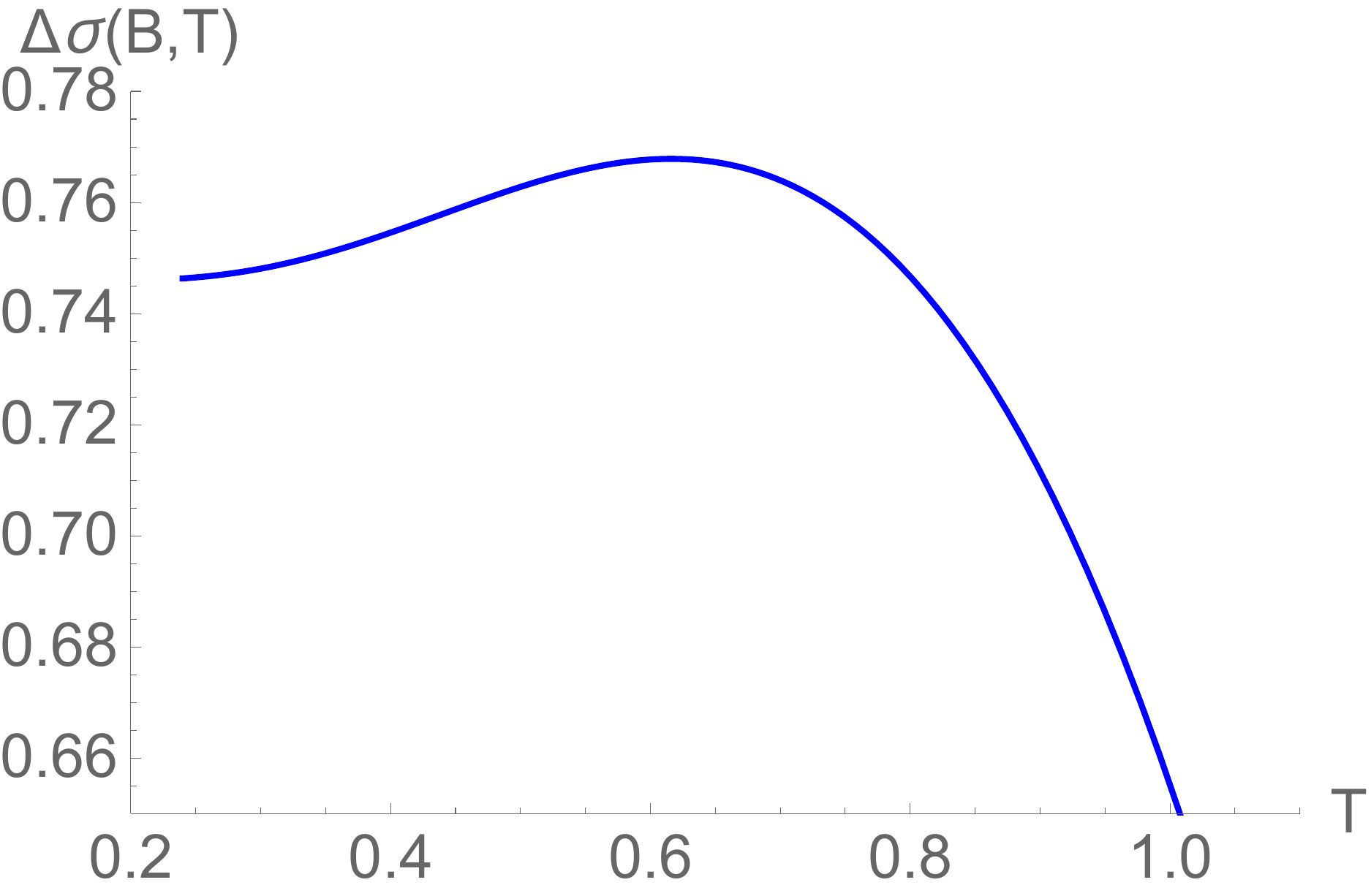}
\caption{\small $\Delta\sigma(B,T)=\sigma(B,T)-\sigma(B=0,T=0)$ as a function of temperature $T$ near the inflection point in the deconfined phase for the case $A_1(z)=-a z^2$. Here $B=0.3$ and a quark mass $m_q=1.0$~\text{GeV} are used. In units \text{GeV}.}
\label{TvsSigmaBPt3Mq1deconfinedalpha1case1}
\end{minipage}
\end{figure}

Let us first discuss the numerical results for the chiral condensate in the confined phase for different values of $B$. For this purpose, we use the thermal-AdS background. The numerical result for $m_q=1.0$~GeV is shown in Figure~\ref{BvsSigmaMq1confinedalpha1case1}. We find that the magnitude of $\sigma$ increases with $B$. This suggests magnetic catalysis behaviour in the confined phase. This result also collaborates well with lattice QCD findings, where a similar magnetic catalysis behaviour has been observed in the confined phase \cite{Bali:2012zg}.

Our numerical results for the thermal profile of the quark condensate in the deconfinement phase is shown in Figure~\ref{TvsSigmavsBMq1deconfinedalpha1case1}. We find that for a fixed value of $B$ the magnitude of $\sigma$ first increases and then decreases with temperature. This behaviour of the chiral condensate can be compared with the lattice findings, where a similar non-monotonic behaviour in the chiral condensate was observed \cite{Bali:2011qj,Bali:2012zg}. The low temperature non-monotonic behaviour of the condensate is explicitly demonstrated in Figure~\ref{TvsSigmaBPt3Mq1deconfinedalpha1case1}. Therefore, like in lattice QCD, there exists a inflection point where the curvature of $\sigma$ changes its sign with temperature, i.e.~(convex $\leftrightarrow$ concave). Notice that, since the temperature dependence of $\braket{\bar{\psi}\psi}$ comes only from $\sigma\propto m_q$, the inflection points of $\braket{\bar{\psi}\psi}$ and $\sigma$ coincide and moreover, they will be $m_q$-independent. Following the lattice work \cite{Bali:2011qj}, we can therefore also define the chiral critical temperature $T_{chiral}$ via this inflection point. A similar definition has also been adopted in other holographic related work, see \cite{Ballon-Bayona:2020xtf}. We can, up to a sign convention, compare our curve with \cite[bottom left of Fig.~5]{Bali:2011qj}. One noteworthy difference is that our $\Delta\sigma(B,T)$ does not saturate near zero for larger temperatures, this is again rooted in having the linear equation \eqref{chiraleq}, that is, not having a potential for $X$ present.

\begin{figure}[h!]
\begin{minipage}[b]{0.5\linewidth}
\centering
\includegraphics[width=2.8in,height=2.0in]{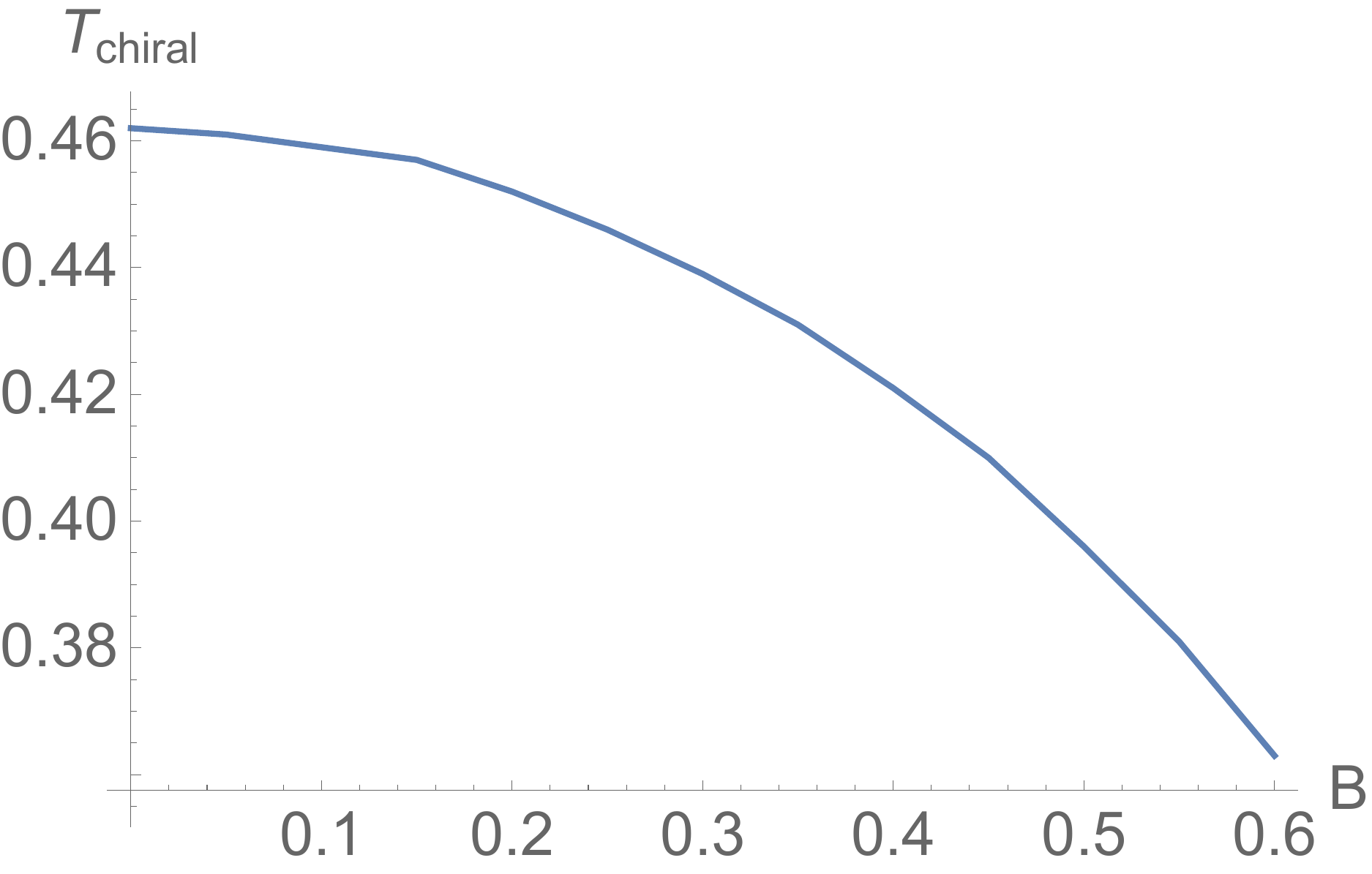}
\caption{ \small Variation of the chiral critical temperature with respect to the magnetic field $B$ for the case $A_1(z)=-a z^2$. Here $m_q=1.0$ is used. In units \text{GeV}.}
\label{BvsTchiralMq1withdilatoncase1}
\end{minipage}
\hspace{0.4cm}
\begin{minipage}[b]{0.5\linewidth}
\centering
\includegraphics[width=2.8in,height=2.3in]{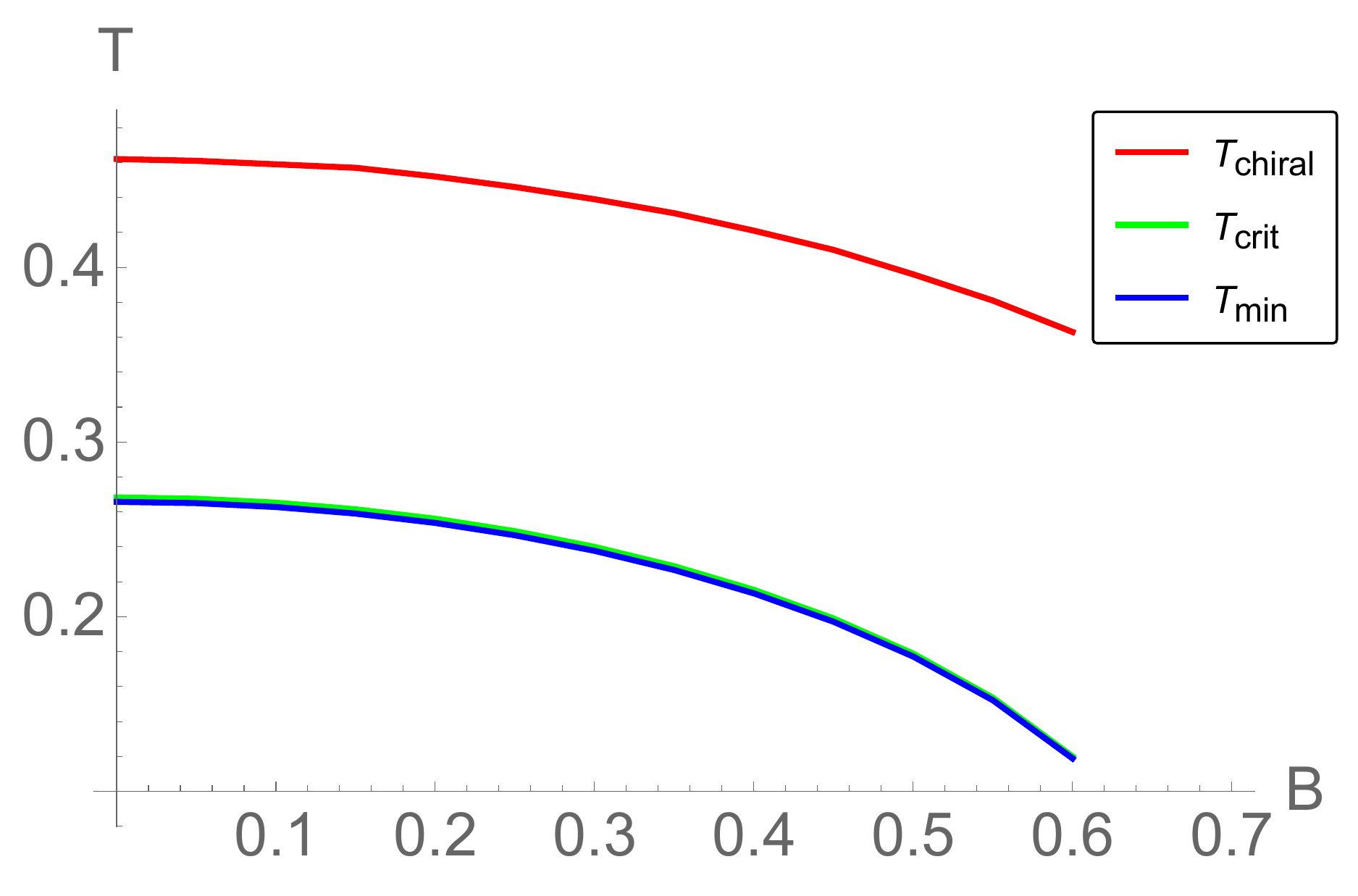}
\caption{\small Variation of the chiral critical temperature, deconfinement critical temperature and $T_{min}$ with respect to the magnetic field $B$ for the case $A_1(z)=-a z^2$. In units \text{GeV}.}
\label{BvsdifferentTcritMq1case1}
\end{minipage}
\end{figure}

In Figure~\ref{BvsTchiralMq1withdilatoncase1}, we have plotted the chiral critical temperature, computed from the inflection point, with respect to the magnetic field. To find the critical temperature, we used a high degree polynomial fit for all data around the inflection point and used that fit to estimate to a high degree of precision a sign change in the second derivative of $\sigma$.  We find that the inflection point goes down with the magnetic field. This is an indication that the model exhibits inverse magnetic catalysis behaviour in the deconfinement phase. This result should be contrasted with the soft wall models, where instead magnetic catalysis behaviour was observed in the chiral critical temperature.

This again indicates the versatility of our holographic model in mimicking lattice QCD properties. Overall, our holographic results for $T_{chiral}(B)$ agree qualitatively well with state-of-the-art lattice results \cite{Bali:2011qj,Bali:2012zg}, albeit with a slightly larger magnitude. For completeness, the magnetic field dependence of the deconfinement and chiral critical temperature, along with $T_{min}$, \footnote{$T_{min}$ is the minimum temperature below which black hole solution does not exist. See \cite{Bohra:2019ebj} for more details.} are combined in Figure~\ref{BvsdifferentTcritMq1case1}.

Note also that in most AdS/QCD models the black hole geometry is thermodynamically stable only above the Hawking-Page critical temperature, i.e.~it only applies at $T > T_{crit}(B)$. Accordingly, one can study the thermal profile of the quark condensate based on the black hole metric only above $T_{crit}(B)$, i.e.~in the deconfined phase. In the confined phase (dual to thermal-AdS), the temperature does not appear in the geometry itself. This implies that the chiral condensate would be a temperature independent constant all the way up to $T_{crit}(B)$ and afterwards, it would follow an analogous condensate pattern for $T> T_{crit}(B)$ as shown before. A troublesome side effect of this is that the chiral condensate would generally exhibit a discontinuous jump at $T_{crit}(B)$, which looks problematic. In particular, the discontinuous jump of the chiral condensate at $T_{crit}(B)$ is in contrast with lattice findings, where no such jump is observed. Unfortunately, this is an inherent property of most holographic QCD models (arising because of implicit $N\to\infty$ approximation). This was already established in \cite{Colangelo:2011sr}.  Interestingly, thereby giving further credit to our model, the magnitude of the chiral condensate at $T_{crit}$ and at $T=0$ almost coincide in our model, i.e.~$\sigma(B,T=0)\simeq \sigma(B,T=T_{crit})$.  Moreover, this result remains true for the second $A_2(z)$ form factor as well. This indicates, yet again, from a different perspective that our holographic model performs reasonably well in mimicking real QCD properties.

\subsection{Using the form factor $A(z)=A_2(z)=-a z^2 - d\, B^2 z^5$}
\begin{figure}[h!]
\centering
\includegraphics[width=2.8in,height=2.3in]{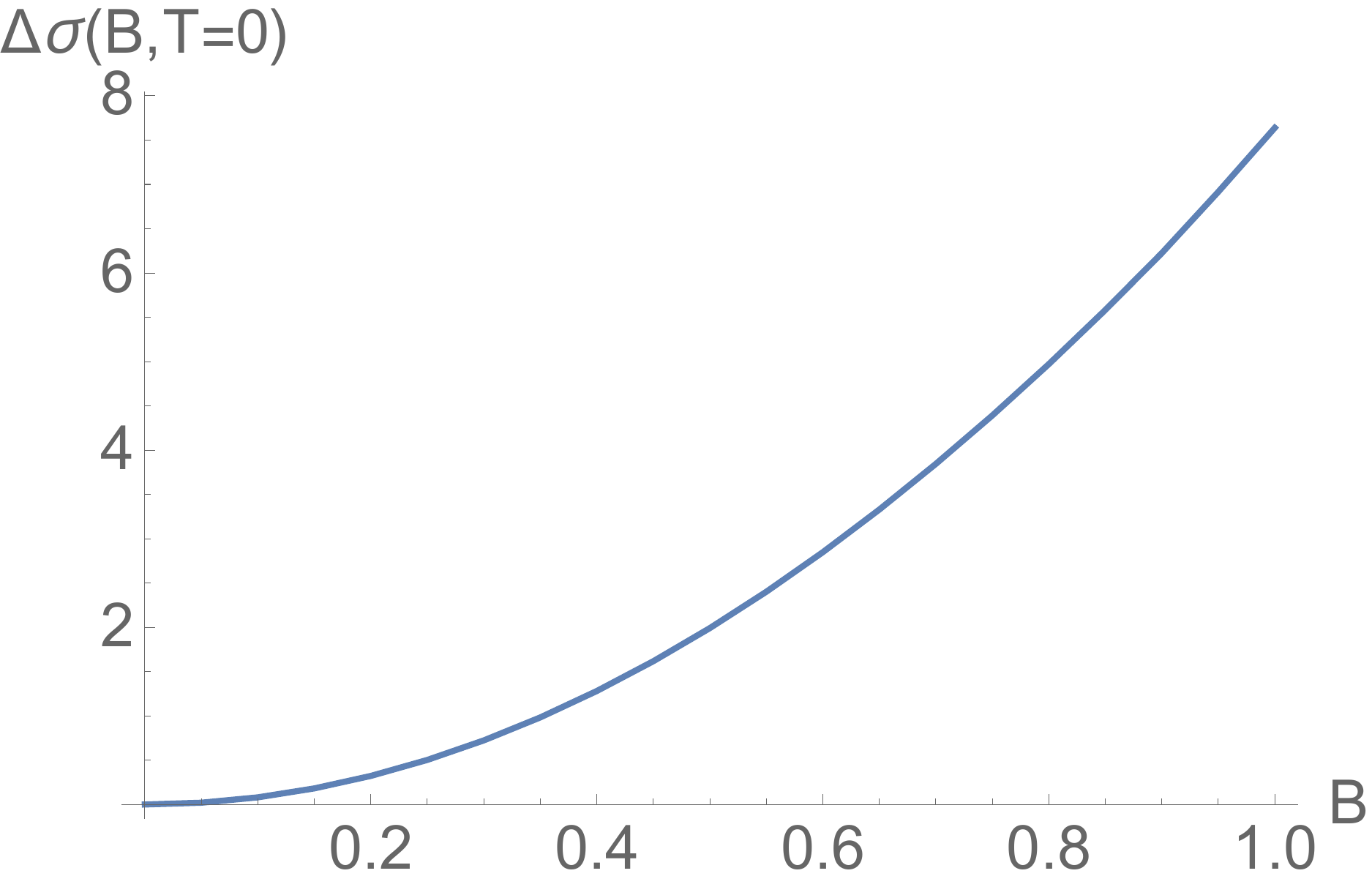}
\caption{\small $\Delta\sigma(B,T=0)=\sigma(B,T=0)-\sigma(B=0,T=0)$ as a function of magnetic field $B$ in the confined phase for the case $A_2(z)=-a z^2-d B^2 z^5$.  Here a quark mass $m_q=1.0$ is used. In units of \text{GeV}.}
\label{BvsSigmaMq1confinedalpha1case2}
\end{figure}
\begin{figure}[h!]
\begin{minipage}[b]{0.5\linewidth}
\centering
\includegraphics[width=2.8in,height=2.0in]{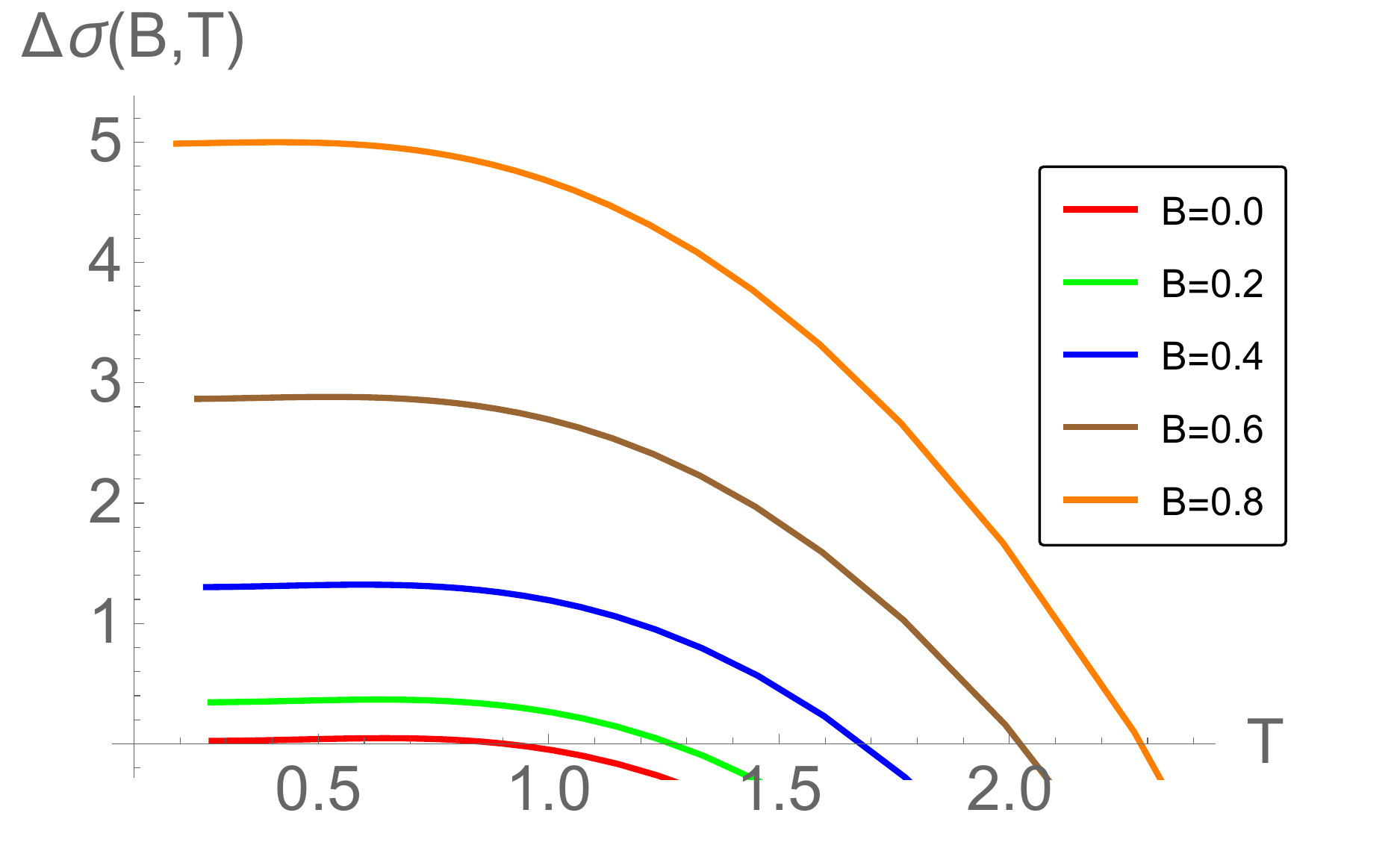}
\caption{ \small $\Delta\sigma(B,T)=\sigma(B,T)-\sigma(B=0,T=0)$ as a function of temperature $T$ in the deconfined phase for the case $A_2(z)=-a z^2-d B^2 z^5$ for different values of the magnetic field.  Here a quark mass $m_q=1.0$ is used. In units \text{GeV}.}
\label{TvsSigmavsBMq1deconfinedalpha1case2}
\end{minipage}
\hspace{0.4cm}
\begin{minipage}[b]{0.5\linewidth}
\centering
\includegraphics[width=2.8in,height=2.3in]{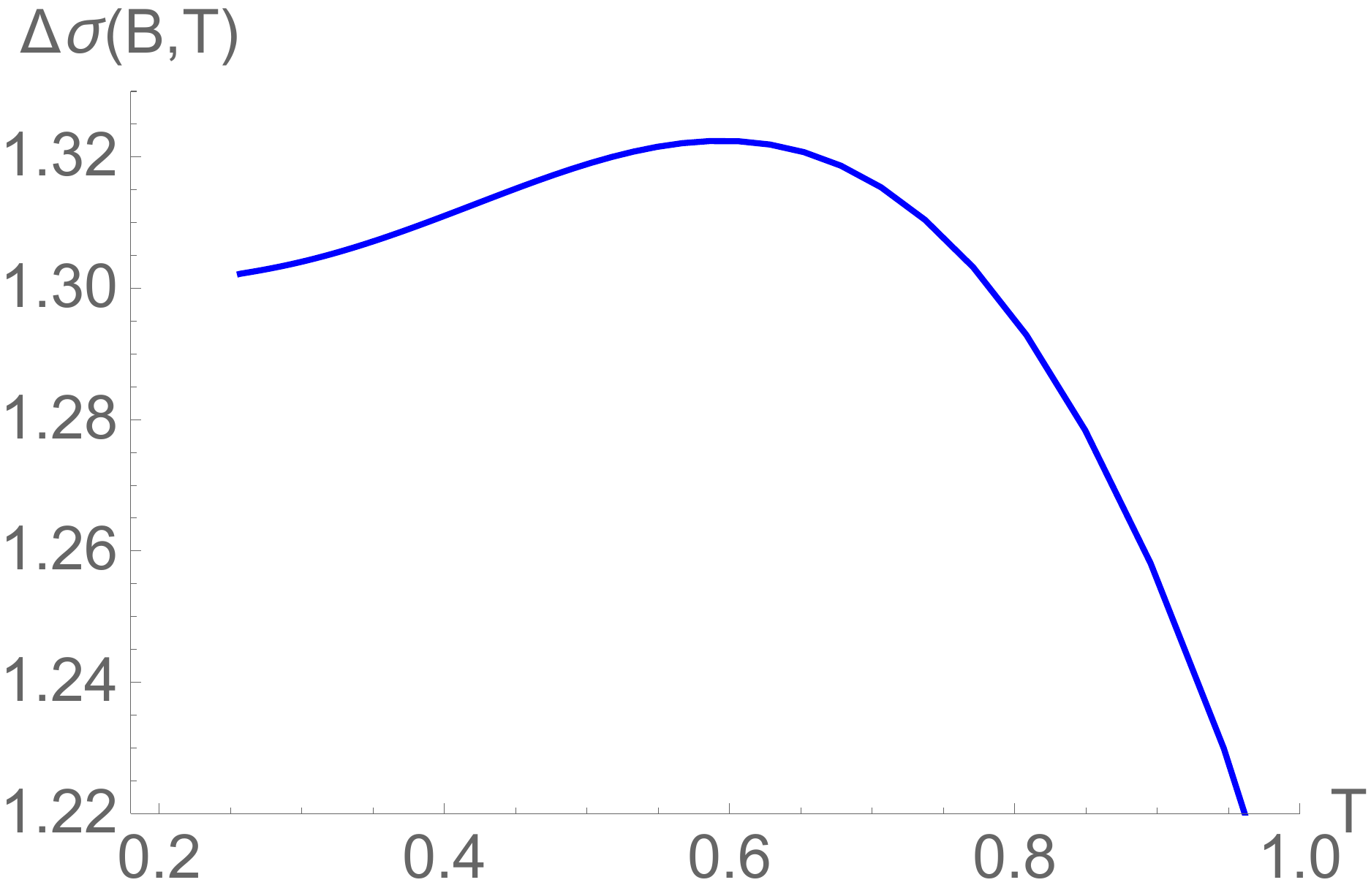}
\caption{\small The profile of $\Delta\sigma(B,T)=\sigma(B,T)-\sigma(B=0,T=0)$ as a function of temperature $T$ near the inflection point in the deconfined phase for the case $A_2(z)=-a z^2-d B^2 z^5$. Here $B=0.4$ and a quark mass $m_q=1.0$ are used. In units \text{GeV}.}
\label{TvsSigmaBPt4Mq1deconfinedalpha1case2}
\end{minipage}
\end{figure}
\begin{figure}[h!]
\begin{minipage}[b]{0.5\linewidth}
\centering
\includegraphics[width=2.8in,height=2.0in]{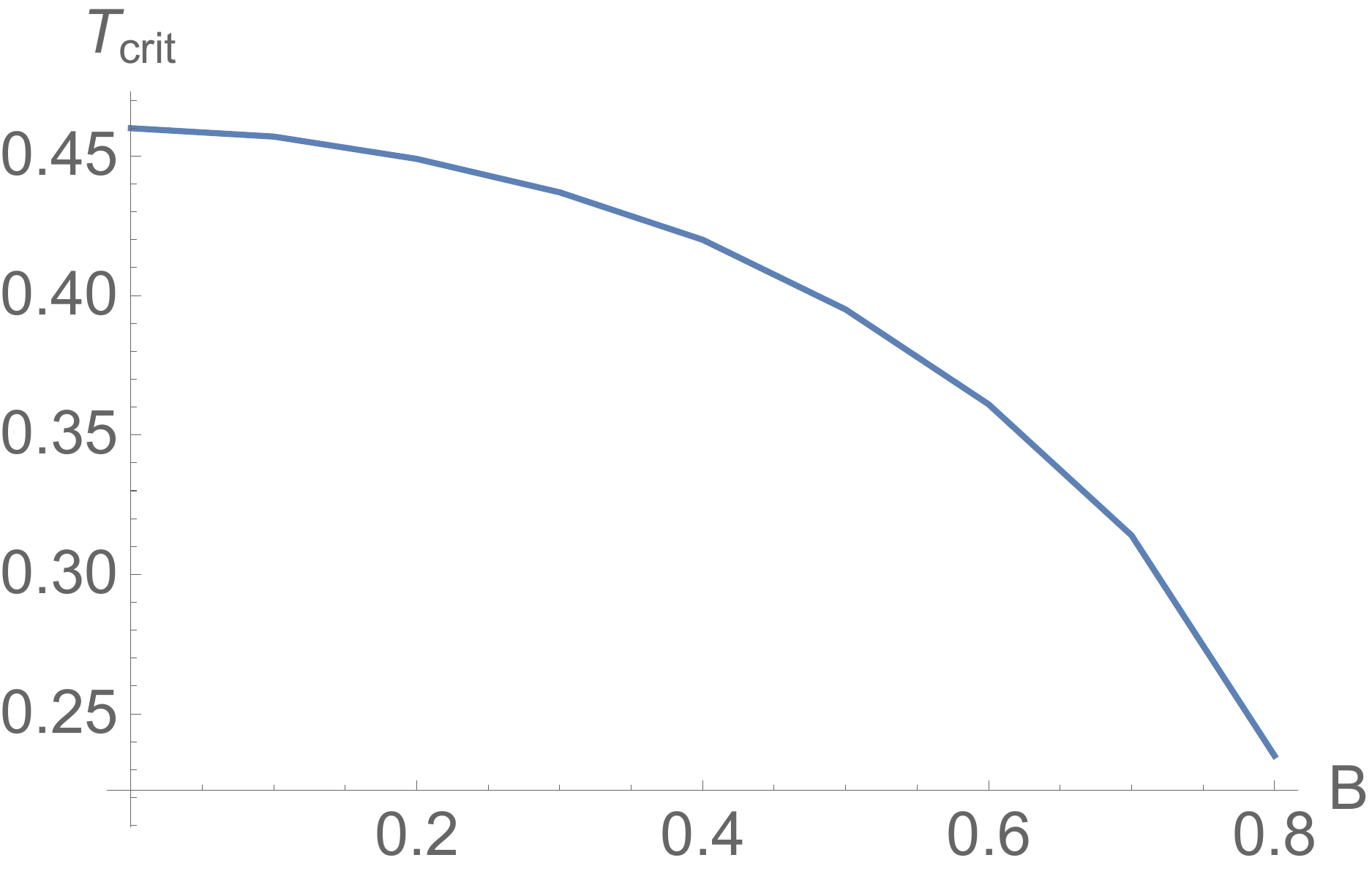}
\caption{ \small Variation of the chiral critical temperature with respect to the magnetic field $B$ for the case $A_2(z)=-a z^2-d B^2 z^5$. Here $m_q=1.0$ is used. In units \text{GeV}.}
\label{BvsTchiralMq1withdilatoncase2}
\end{minipage}
\hspace{0.4cm}
\begin{minipage}[b]{0.5\linewidth}
\centering
\includegraphics[width=2.8in,height=2.3in]{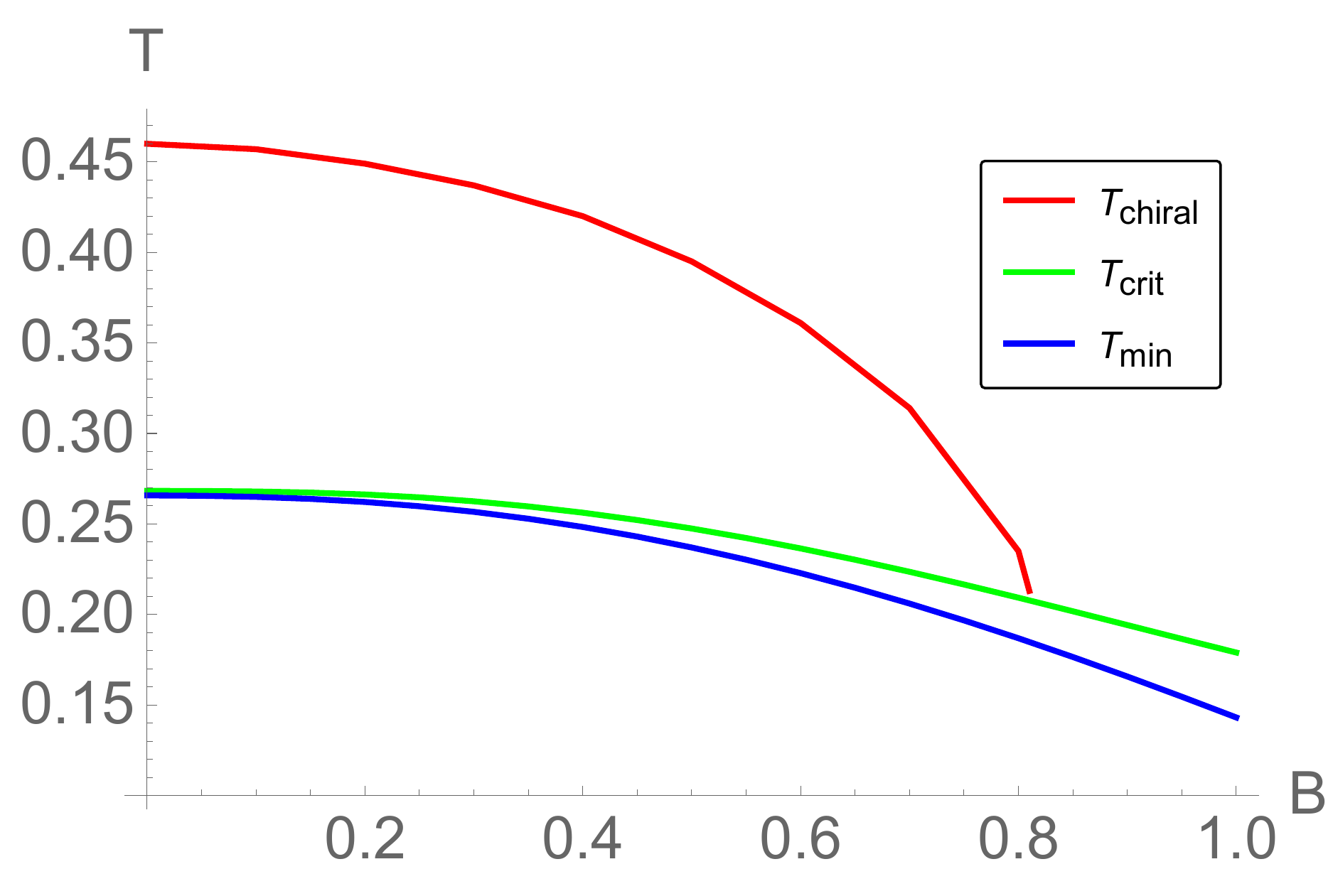}
\caption{\small Variation of the chiral critical temperature, deconfinement critical temperature and $T_{min}$ with respect to the magnetic field $B$ for the case $A_2(z)=-a z^2-d B^2 z^5$. In units \text{GeV}.}
\label{BvsdifferentTcritMq1case2}
\end{minipage}
\end{figure}

Let us now discuss the chiral sector of the dual boundary theory using the second form factor $A(z)=A_2(z)$. Most of the numerical routine and procedure are similar to what we alluded to in the above subsection,  and can be straightforwardly generalised to the $A_2(z)$ case. The near boundary expansion of the chiral field $X$ remains the same (cf.~eq.~(\ref{Xnearzhexp})) whereas the near horizon expansion changes accordingly (eq.~(\ref{Xnearzhhor1})). Since most of the chiral analysis is similar to the previous case, therefore, we can be rather brief here.

Our results for the chiral condensate for $A_2(z)$ in the confined phase is shown in Figure~\ref{BvsSigmaMq1confinedalpha1case2}. As the allowed range of the magnetic field increases for $A_2(z)$, we can now probe the chiral condensate for slightly larger magnetic field values. For this form factor as well, the chiral condensate is found to be increasing with the magnetic field. This again indicates magnetic catalysis behaviour in the chiral sector of the confined phase, collaborating once again qualitatively well with lattice results.

The thermal profile of the condensate results in the deconfinement phase is shown in Figure~\ref{TvsSigmavsBMq1deconfinedalpha1case2}. The condensate again first increases and then decreases with temperature, therefore exhibiting a lattice like non-monotonic thermal behaviour. The low temperature near-inflection behaviour of the condensate is shown in Figure~\ref{TvsSigmaBPt4Mq1deconfinedalpha1case2}. Therefore, there exists an inflection temperature where the curvature of $\sigma$ changes from convex to concave profile and vice versa. Our numerical analysis suggests that this inflection point exists only up to $B \simeq 0.8~\text{GeV}$, and for higher magnetic field values it ceases to exist. To be more precise, the inflection point is moved below the deconfinement temperature $T_{crit}(B)$ where the black hole metric no longer applies.  The chiral critical temperature, computed from the inflection point, with respect to the magnetic field is shown in Figure~\ref{BvsTchiralMq1withdilatoncase2}. It again comes out to be a decreasing function of the magnetic field, thereby explicitly confirming the inverse magnetic catalysis behaviour in the chiral sector. Importantly, the dual boundary theory exhibits inverse magnetic catalysis behaviour for all values of $m_q$. Further, in this case as well, the magnitude of the chiral condensate at $T_{crit}$ and at $T=0$ almost coincide, i.e.~$\sigma(B,T=0)\simeq \sigma(B,T=T_{crit})$, which we interpret as a good sign. In Figure~\ref{BvsdifferentTcritMq1case2}, the relative magnitude of the chiral critical, deconfinement critical and $T_{min}$ temperatures are shown for completeness. For this form factor as well, the chiral critical temperature turns out to be slightly higher than the deconfinement critical temperature.

\section{Outlook}
We have continued the research set out in \cite{Bohra:2019ebj} and constructed an Einstein-Maxwell-dilaton gravity model that not only captures the inverse magnetic catalysis for the deconfinement transition, but also the same phenomenon affecting the chiral transition.  As a byproduct, we also investigated the anisotropy in the string tension, showing that linear confinement is always realized within the range of validity of our model. These findings are in qualitative agreement with other studies, in particular those coming from lattice simulations.

We computed the chiral condensate in the confined and deconfined phases using two form factors and found form-independent chiral features. The condensate magnitude was found to be increasing with $B$ in the confined phase, thereby suggesting magnetic catalysis, whereas the inflection temperature in the deconfined phase was found to be decreasing with $B$, thereby suggesting inverse magnetic catalysis in the chiral critical temperature. These results agree qualitatively well with lattice results. However, unfortunately, $T_{chiral}(B)$ comes out to be slightly larger compared to lattice results. We have investigated a few other relatively simple form factors as well and found similar results in all these cases. It thus appears that inverse catalysis behaviour in the chiral sector is a generic feature of our model. To improve our model further, we plan on including a potential $V_C(X)$ for the chiral $X$-field, following earlier efforts as in \cite{Gherghetta:2009ac,Chen:2019rez}, the latter without magnetic field though. This will make the chiral condensate a truly dynamical feature, as for now, the crucial parameter $\sigma$ entering the chiral condensate \eqref{psibarpsi2} is directly proportional to the bare quark mass $m_q$, as can be rapidly proven from the boundary expansion \eqref{Xnearzhexp}.

A subtle point in such construction will also be the correct identification of the chiral condensate by extending the analysis in our Appendix B, as \cite{Colangelo:2011sr,Gherghetta:2009ac,Chen:2019rez} considered the $\sigma$ entering the boundary expansion of $X$ as an avatar for the chiral condensate, but a correct identification is more subtle, in particular at finite $B$, as we have shown, see also \cite{Dudal:2015wfn}.

Another issue worthy of our attention will be the proper identification of the boundary magnetic field in terms of the bulk one. To do this properly relative to QCD, we should use $N_f=2$ flavours in the chiral sector and mimic a magnetic field by slightly gauging the unbroken diagonal sector of the underlying $U(1)_b\times SU(2)_V \times SU(2)_C$ model \cite{Sakai,Callebaut:2013wba}, rather than the current simplified $U(1)\times U(1)$ version. $U(1)_b$ refers to the baryon number current, $SU(2)_V$ to the (unbroken) flavour symmetries and $SU(2)_C$ to the (broken) chiral symmetries, with $SU(2)_L\times  SU(2)_R \simeq SU(2)_V\times SU(2)_C$. This also implies that there will be a more direct link between the chiral and EMD-action, also requiring a proper study of the coupling prefactors of both parts of the action in relation to QCD OPE results.

In principle, we should also try to include the back reaction of the chiral field $X$ into the Einstein equations of motions, which would correspond the unquenching of our setup. Although this might sound as an ambitious step, it might be possible via a generalization of the potential reconstruction method, in combination with a phenomenological profile for the potential $V_C(X)$ as proposed in e.g.~\cite{Gherghetta:2009ac}.

Once all (or at least most) of the above is achieved, we can aim at studying magnetic field dependent QCD observables that are not accessible via lattice simulations, such as various transport properties.

We will report on these and other topics in the near future.

\section*{Acknowledgments}
A.H.~would like to thank a scholarship that has been awarded by the Ministry of Science, Research and
Technology (Department of Scholarship and Students' Affairs Abroad) of the Islamic Republic of Iran
which made the initial stages of this research at KU Leuven--Kulak possible. The work of S.M.~is supported by the Department of Science and
Technology, Government of India under the Grant Agreement number IFA17-PH207 (INSPIRE Faculty
Award).

\appendix

\section{Appendix A: (In)dependence of the potential on temperature and magnetic field for the new form factor $A_2(z)$}
\begin{figure}[h!]
\begin{minipage}[b]{0.5\linewidth}
\centering
\includegraphics[width=2.8in,height=2.3in]{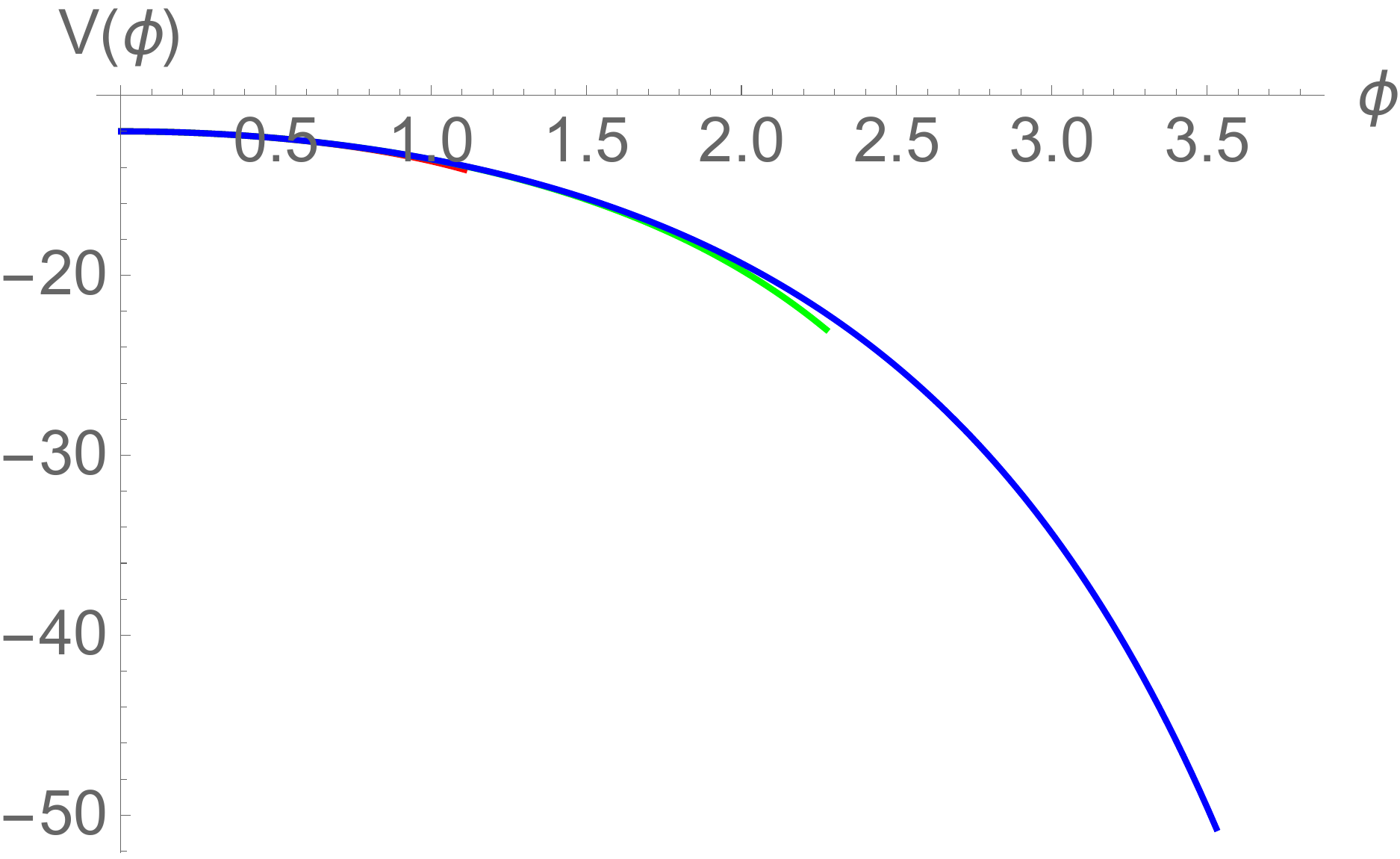}
\caption{ \small The variation of potential as a function of $\phi$ for different $z_h$. Here
and $B=0.3$ is considered. Here red, green and blue curves correspond to $z_h=0.5$, $1.0$
and $1.5$ respectively.}
\label{phivsVvszhBpt3case2}
\end{minipage}
\hspace{0.4cm}
\begin{minipage}[b]{0.5\linewidth}
\centering
\includegraphics[width=2.8in,height=2.3in]{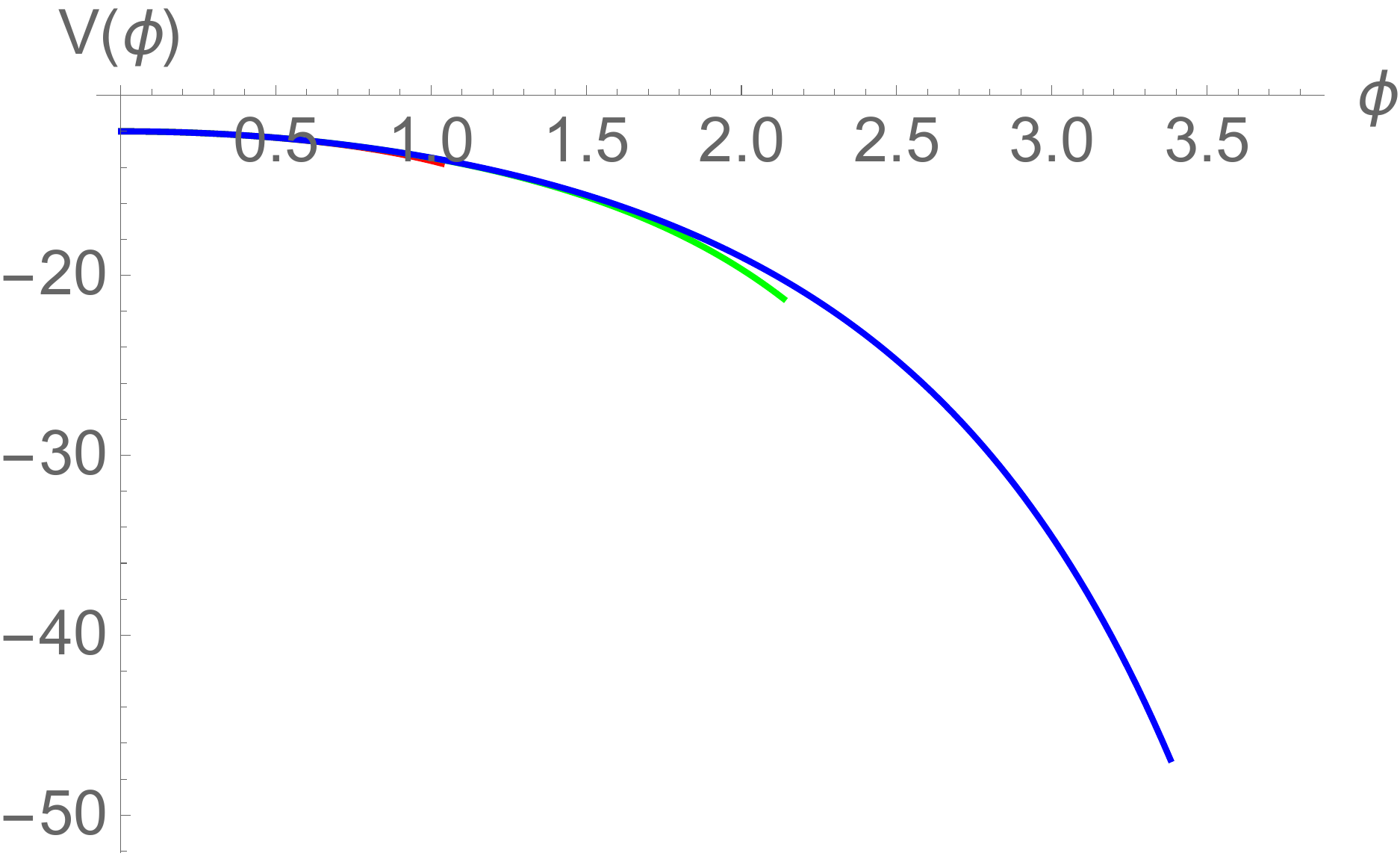}
\caption{\small The variation of potential as a function of $\phi$ for different $z_h$. Here
and $B=0.5$ is considered. Here red, green and blue curves correspond to $z_h=0.5$, $1.0$
and $1.5$ respectively.}
\label{phivsVvszhBpt5case2}
\end{minipage}
\end{figure}

\begin{figure}[h!]
\begin{minipage}[b]{0.5\linewidth}
\centering
\includegraphics[width=2.8in,height=2.3in]{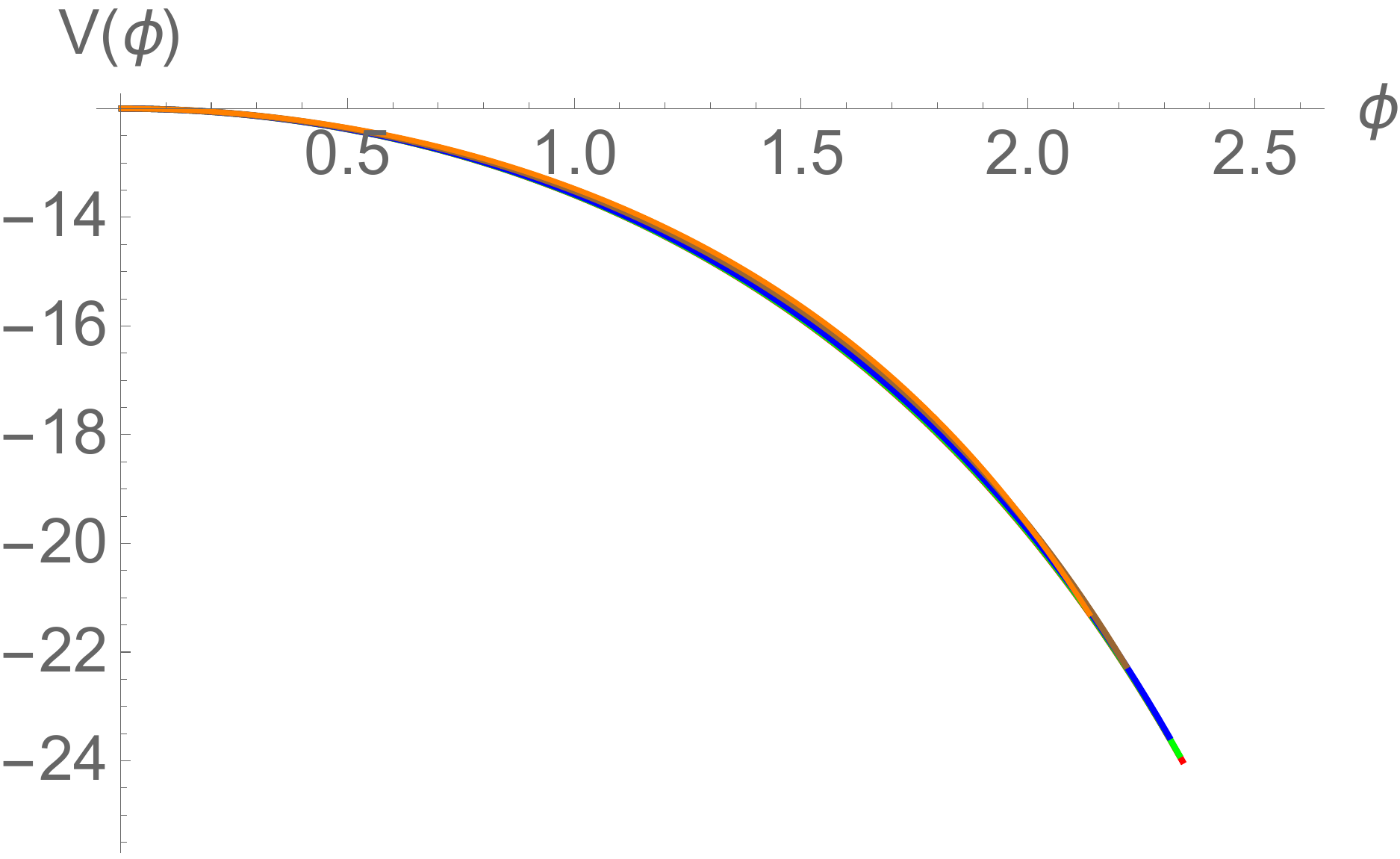}
\caption{ \small The variation of potential as a function of $\phi$ for different $B$. Here
and $z_h=1.0$ is considered. Here red, green, blue, brown and orange curves correspond to $B=0.1$, $0.2$, $0.3$, $0.4$
and $0.5$ respectively.}
\label{phivsVvsBzh1case2}
\end{minipage}
\hspace{0.4cm}
\begin{minipage}[b]{0.5\linewidth}
\centering
\includegraphics[width=2.8in,height=2.3in]{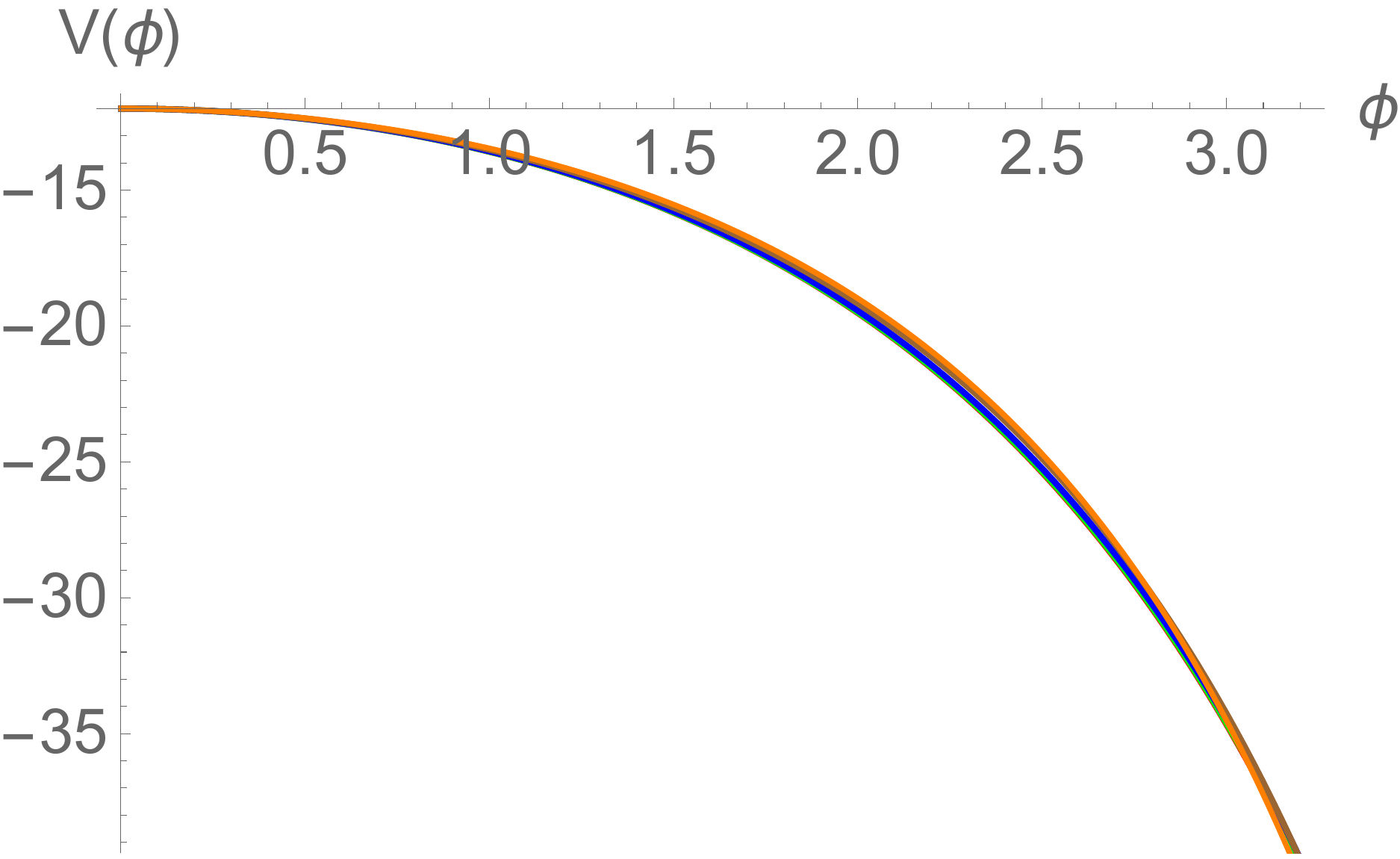}
\caption{\small The variation of potential as a function of $\phi$ for different $B$. Here
and $z_h=1.5$ is considered. Here red, green, blue, brown and orange curves correspond to $B=0.1$, $0.2$, $0.3$, $0.4$
and $0.5$ respectively.}
\label{phivsVvsBzh1pt5case2}
\end{minipage}
\end{figure}

In Figures~\ref{phivsVvszhBpt3case2}-\ref{phivsVvsBzh1pt5case2}, the almost independent behaviour of the dilaton potential on $T$ and $B$ for the new form factor $A_2(z)=-a z^2 - d\, B^2 z^5$ is shown. A similar analysis was performed in \cite{Bohra:2019ebj} for the first form factor $A_1(z)=-a z^2$.

\section{Appendix B: the chiral condensate}
In this Appendix, we derive the relation between the coefficient $\sigma$ and the chiral condensate $\braket{\bar{\psi}\psi}$, following \cite{Dudal:2015wfn}. From the Lagrangian $\mathcal{L}=\bar{\psi}(\gamma^\mu \partial_\mu-m_q)\psi$, the $\braket{\bar{\psi}\psi}$ condensate can be obtained by differentiating the partition function $W=\ln Z $ with respect to $m_q$ as,
\begin{eqnarray}
\frac{1}{Z}\frac{dZ}{dm_q} = \frac{\int [\mathcal{D}\psi \mathcal{D}\bar{\psi}] \  (\int d^4 x \bar{\psi}\psi) \ e^{-\int d^4 x \mathcal{L}}}{\int [\mathcal{D}\psi \mathcal{D}\bar{\psi}] \ e^{-\int d^4 x \mathcal{L}}} \,.
\end{eqnarray}
Here we restrict ourselves to the single quark flavour sector. Using the gauge-gravity duality and equating the bulk and the boundary partition functions, i.e.~$Z=e^{-S_{chiral}}$, we obtain
\begin{eqnarray}
V_4 \braket{\bar{\psi}\psi} = -\frac{d}{dm_q}\left(\frac{N_c}{16 \pi^2}\int d^5x \sqrt{-g} e^{-\phi} \ \left[\partial_\mu X \partial^\mu X + m_{5}^2 X^2\right]       \right)
\end{eqnarray}
where $V_4$ is the volume of the four-dimensional boundary spacetime. Using the $X$-equation of motion and restricting to a homogeneous condensate $X(z,x^\mu)=X(z)$, we can further simplify the above expression to
\begin{eqnarray}
& & \braket{\bar{\psi}\psi} = -\frac{N_c}{16 \pi^2} \frac{d}{dm_q}\left( \sqrt{-g} e^{-\phi} g^{zz} X(z)X'(z)\bigg\rvert_{z=0}^{z=z_h} \right)  =  \frac{N_c}{16 \pi^2} \frac{d}{dm_q}\left( \sqrt{-g} e^{-\phi} g^{zz} X(z)X'(z)\bigg\rvert_{z=0}\right) \,.
\label{psibarpsi}
\end{eqnarray}
Here we have considered the black hole background and have used the fact that $g^{zz}(z_h)=0$. For the thermal-AdS background, the upper limit $z=z_h$ in the above equation will get replaced by $z=\infty$. Substituting the near boundary expansion of the $X$-field,
\begin{eqnarray}
X(z)= m_q z + m_q b_1  z^2 +  \sigma z^3 +  m_q b_2 z^3 \ln{\sqrt{a}z} + \mathcal{O}(z^4) \,,
\end{eqnarray}
into eq.~(\ref{psibarpsi}) and simplifying, we get
\begin{eqnarray}
& & \braket{\bar{\psi}\psi} = \frac{N_c}{8 \pi^2} \left(\frac{m_q}{\epsilon^2}  - \frac{8 m_q \sqrt{9a-B^2}}{\epsilon} + 4 m_q b_2 \log{\sqrt{a}\epsilon} + 4 \sigma + b_2 m_q -21 B^2 m_q +195 a m_q \right) \,.
\label{psibarpsi1}
\end{eqnarray}
Using the minimal subtraction scheme and removing the divergent terms by hand, we get desired equation,
\begin{eqnarray}
\braket{\bar{\psi}\psi}_{B,T} = \frac{N_c}{2 \pi^2} \sigma(B,T)+ \frac{N_c m_q}{8  \pi^2} \biggl( - 18 B^2 + 165 a \biggr) \,.
\label{psibarpsi2}
\end{eqnarray}

\section{Appendix C: a few words about the chiral action }
Generally in the probe soft wall models, a multiplicative dilaton factor $e^{-\phi}$  is commonly included in the chiral action in an ad-hoc way, this to get well defined QCD-like properties holographically \footnote{In these soft wall models, the factor $e^{-\phi}$ was introduced merely based on the ``analogy'' with probe matter coming from branes. At the practical level, the dilaton prefactor $e^{-\phi}$ was considered to ``smoothly cut-off'' the standard AdS geometry to ensure confinement dynamics.} \cite{Colangelo:2011sr,Dudal:2015wfn}. The main reason for this is that the usual background geometry does not contain the back reaction of the dilaton field and, therefore, the thermal-AdS geometry does not really correspond to the confined phase. Moreover, in soft wall models, the confinement-deconfinement phase transition exists only because of this additional dilaton term. Now, in our model, the back reaction of the dilaton field is introduced consistently from the beginning, and, correspondingly, we have a genuine confinement-deconfinement phase transition. Therefore, we do not necessarily need to include the dilaton field in the chiral action in an ad-hoc manner, as the self-consistent background geometry could takes care of it. Therefore, in our model, we actually have two choices: (i) to include the dilaton factor, and (ii) not to include the dilaton factor. It is hence important to analyse both these choices carefully and see the similarities and differences in their boundary chiral properties. In Sect.~3, we analysed the chiral properties using the first choice by including the dilaton field and found many important chiral features holographically. Here, we will consider the second choice and analyse what would happen if we do not include the dilaton field in the chiral action.

In this case, the relevant chiral action would be,
\begin{eqnarray}
S_{chiral} =  \frac{N_c}{16 \pi^2} \int \mathrm{d^5}x \sqrt{-g} \ \text{Tr}\left[|DX|^2-m_5^2|X|^2-\frac{f_2(\phi)}{3}(F_{L}^2+F_R^2)\right]\,.
\end{eqnarray}
This would lead to $X$-field equation of motion,
\begin{eqnarray}
X''(z)+X'(z)\biggl(-\frac{3}{z}+2 B^2 z + \frac{g'(z)}{g(z)} + 3 A'(z)  \biggr) + \frac{3 e^{2A(z)}X(z)}{z^2 g(z)} =0 \,.
\end{eqnarray}
Following the steps of Sect.~3, we can similarly find UV and IR expansions of $X$ field from which boundary chiral properties can be extracted. In particular, the near UV boundary expansion is
\begin{eqnarray}
X(z)= m_q z + \sigma z^3 + m_q (6a-B^2) z^3 \ln{\sqrt{a}z} + \mathcal{O}(z^4) \,.
\end{eqnarray}
The chiral condensate $\braket{\bar{\psi}\psi}$ is now related to $\sigma$ in the following way,
\begin{eqnarray}
& & \braket{\bar{\psi}\psi} = \frac{N_c}{8 \pi^2} \left(\frac{m_q}{\epsilon^2} + 4 m_q (6a-B^2) \log{\sqrt{a}\epsilon} + 4 \sigma + 3 a m_q \right) \,.
\end{eqnarray}
We have again introduced a UV cut-off $z=\epsilon$. Notice that, compared to the results of Appendix B, there is no $1/\epsilon$ divergent term. Using the minimal subtraction scheme, we can again consider only the finite part, which gives us a relation
\begin{eqnarray}
\braket{\bar{\psi}\psi}_{B,T} = \frac{N_c}{2 \pi^2} \sigma(B,T)+ \frac{3 a N_c m_q}{8 \pi^2} \,.
\label{psibarpsi1}
\end{eqnarray}

\begin{figure}[h!]
\begin{minipage}[b]{0.5\linewidth}
\centering
\includegraphics[width=2.8in,height=2.3in]{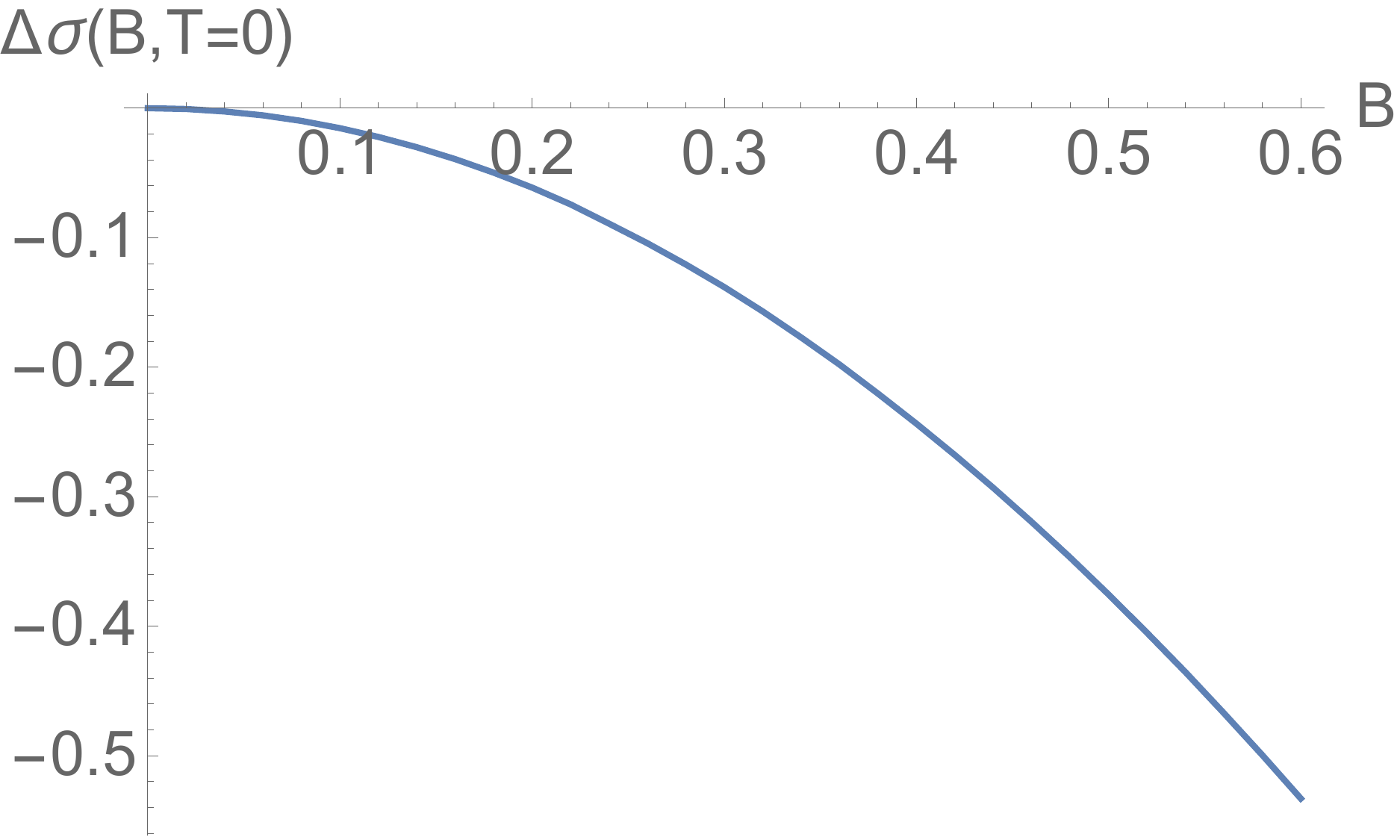}
\caption{ \small $\Delta\sigma(B,T=0)=\sigma(B,T=0)-\sigma(B=0,T=0)$ as a function of magnetic field $B$ in the confined phase for the case $A_1(z)=-a z^2$.  Here a quark mass $m_q=1.0$ is used. In units of ~\text{GeV}.}
\label{BvsSigmaMq1confinedalpha0case1}
\end{minipage}
\hspace{0.4cm}
\begin{minipage}[b]{0.5\linewidth}
\centering
\includegraphics[width=2.8in,height=2.3in]{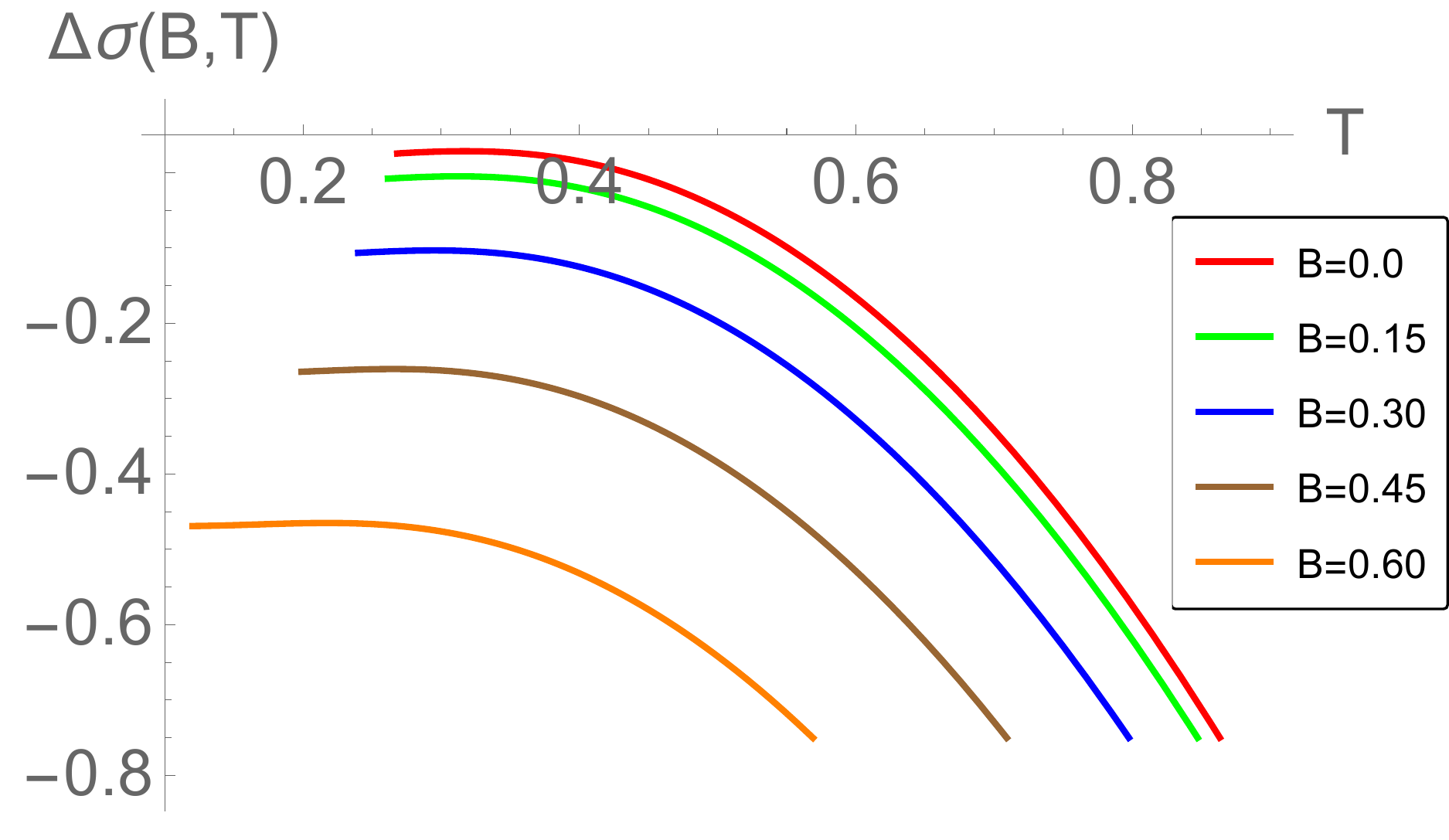}
\caption{\small $\Delta\sigma(B,T)=\sigma(B,T)-\sigma(B=0,T=0)$ as a function of temperature $T$ in the deconfined phase for the case $A_1(z)=-a z^2$ for different values of the magnetic field.  Here a quark mass $m_q=1.0$ is used. In units of ~\text{GeV}.}
\label{TvsSigmavsBMq1deconfinedalpha0case1}
\end{minipage}
\end{figure}

In Figure~\ref{BvsSigmaMq1confinedalpha0case1}, the condensate profile in the confined phases is shown. Clearly, the condensate magnitude decreases with $B$, suggesting inverse magnetic catalysis behaviour. This should be contrasted with the dilaton-included case studied in Sect.~3, where magnetic catalysis was instead found in the confined phase. Since lattice results do suggest magnetic catalysis in the confined phase \cite{Bali:2012zg}, therefore, as far as the condensate behaviour in the confined phase is concerned, the soft wall like chiral action, having a dilaton prefactor in the chiral action, seems to be the appropriate choice.

The thermal profile of the chiral condensate in the deconfinement phase is shown in Figure~\ref{TvsSigmavsBMq1deconfinedalpha0case1}. Again the condensate exhibits a non-monotonic profile, which first increases and then decreases. However, unlike in Sect.~3, now there is no inflection point. In particular, $\sigma''(T)$ is always negative and does not go to zero at any temperature. This again brings out the difference with the dilaton-included case, where maxima and inflection point both existed.  This analysis further highlights the subtle differences that the dilaton factor can introduce in the chiral sector of the holographic models. Therefore, as far as the chiral sector is concerned, the dilaton-included action looks to be more relevant for the holographic studies.

\section{Appendix D: a few words about the shooting method }
Here we briefly describe how to calculate the ``chiral parameter'' $\sigma$ via a shooting method. We start from the confining metric and must integrate the equation of motion from $0$ to infinity. In practice, infinity is replaced by a large number, for our purposes $z=10$ suffices. To solve the ODE~(\ref{chiraleq}) from the boundary $z=0$ to $z\approx \infty$, we use the analytical series solution around $z=0$ to set initial values for both $X(\epsilon)$ and $X'(\epsilon)$ at $0<\epsilon\ll1$. Thence, utilizing the near boundary expansion of $X$, eq.~(\ref{Xnearzhexp}), we have
\begin{eqnarray}
X(\epsilon)&=& m_q \epsilon + \sigma \epsilon^3 + m_q n \epsilon^3 \ln{\sqrt{a}\epsilon} \,, \cr
 X'(\epsilon)&=&3nm_q{\epsilon}^2 \ln(\sqrt{a}\epsilon) +(nm_q+3 \sigma) {\epsilon}^2 + m_q~.
 \label{chiepsilon}
\end{eqnarray}
Here, $\sigma$ enters as the shooting parameter. Its value can be fixed by imposing that the numerical solution for $X(z)$ is normalizable and thus a finite number at infinity as it follows from a large $z$ expansion of the equation \eqref{chiraleq}.  This is achieved as follows:
\begin{enumerate}

\item Find initial values $\sigma_1$ and $\sigma_2$, for which $X_{\sigma_1}[10]<0$ and $X_{\sigma_2}[10]>0$.
\item  Define $\sigma_3=\frac{\sigma_1+\sigma_2}{2}$.

\item If $X_{\sigma_3}[10]<0$, then we know the $\sigma$ that we are looking for is located in between $\sigma_2$ and $\sigma_3$.

\item Redefine $\sigma_1=\sigma_3$, and keep $\sigma_2$. Otherwise, if $X_{\sigma_3}[10]>0$, $\sigma_2=\sigma_3$ and keep  $\sigma_1$.

\item Repeat until $|\sigma_1-\sigma_2|<$~preset tolerance level.

\end{enumerate}
Evidently, this method is based on the intermediate value theorem.

\end{document}